\documentclass[]{aastex631}

\usepackage{amsmath}
\usepackage{placeins}
\usepackage{multirow}
\usepackage{booktabs}
\usepackage{txfonts}
\usepackage{rotating}
\graphicspath{{./}{figures/}}
\begin{document}

\title{{\em B}-field Orion Protostellar Survey (BOPS). IV: The Relative Orientation Between Magnetic Fields and Density Structures in Young Protostellar Envelopes}

\author[0009-0003-6377-9023]{Kexin Cai}
\affiliation{Guangxi Key Laboratory for Relativistic Astrophysics, School of Physical Science and Technology, Guangxi University, Nanning 530004, China}

\author[0000-0001-7393-8583]{Bo Huang}
\altaffiliation{Corresponding author}
\email{hb6170@gmail.com}
\affiliation{Institut de Ciències de l'Espai (ICE-CSIC), Campus UAB, Can Magrans S/N, E-08193 Cerdanyola del Vallès, Catalonia, Spain}
\affiliation{Korea Astronomy and Space Science Institute, 776 Daedeok-daero, Yuseong-gu, Daejeon 34055, Republic of Korea}

\author[0000-0002-3829-5591]{Josep M. Girart}
\affiliation{Institut de Ciències de l'Espai (ICE-CSIC), Campus UAB, Can Magrans S/N, E-08193 Cerdanyola del Vallès, Catalonia, Spain}
\affiliation{Institut d’Estudis Espacials de Catalunya (IEEC), Campus del Baix Llobregat - UPC, Esteve Terradas 1, E-08860 Castelldefels, Catalonia, Spain}

\author[0000-0003-3017-4418]{Ian W. Stephens}
\affiliation{Department of Earth, Environment, and Physics, Worcester State University, Worcester, MA 01602, USA}

\author[0000-0002-3078-9482]{\'Alvaro S\'anchez-Monge}
\affiliation{Institut de Ciències de l'Espai (ICE-CSIC), Campus UAB, Can Magrans S/N, E-08193 Cerdanyola del Vallès, Catalonia, Spain}
\affiliation{Institut d'Estudis Espacials de Catalunya (IEEC), c/Gran Capita, 2-4, E-08034 Barcelona, Catalonia, Spain}

\author[0000-0002-5714-799X]{Valentin J. M. Le Gouellec}
\affiliation{Institut de Cienci\`es de l’Espai (ICE-CSIC), Campus UAB, Carrer de Can Magrans S/N, E-08193 Cerdanyola del Vall\`es, Spain}
\affiliation{Institut d’Estudis Espacials de Catalunya (IEEC), c/ Gran Capitá, 2-4, 08034 Barcelona, Spain}

\author[0000-0001-9822-7817]{Wenyu Jiao}
\affiliation{Shanghai Astronomical Observatory, Chinese Academy of Sciences, No.80 Nandan Road, Shanghai 200030, People’s Republic of China}

\author[0000-0002-5077-9599]{Qianru He}
\affiliation{Purple Mountain Observatory, Chinese Academy of Sciences, Nanjing, 210023, China}
\affiliation{School of Astronomy and Space Sciences, University of Science and Technology of China, Hefei, 230026, China}

\author[0000-0002-9138-5940]{Zu-Jia Lu}
\altaffiliation{Corresponding author}
\email{luzujia@gxu.edu.cn}
\affiliation{Guangxi Key Laboratory for Relativistic Astrophysics, School of Physical Science and Technology, Guangxi University, Nanning 530004, China}
\affiliation{Department of Physics, University of Oxford, Keble Road, Oxford OX1 3RH, UK}

\author{Enwei Liang}
\affiliation{Guangxi Key Laboratory for Relativistic Astrophysics, School of Physical Science and Technology, Guangxi University, Nanning 530004, China}

\begin{abstract}
We investigate the relative alignment between density structures and magnetic fields in eight young protostars from the ALMA {\em B}-field Orion Protostellar Survey.
Column density maps are derived from 870~$\mu$m dust continuum emission, and the Histogram of Relative Orientations (HRO) method is applied to quantify the correlation between magnetic field orientations and density structures on envelope scales ($\sim$10$^{3}$~au).
We find that the relative alignment shows overall weak evidence of systematic evolution with column density, suggesting that column density alone does not fully determine the alignment.
The magnetization level also plays a crucial role, with weakly magnetized envelopes exhibiting predominantly parallel or random alignment, whereas strongly magnetized ones show perpendicular configurations even at moderate densities.
These results reveal that density and magnetization jointly shape the morphology of protostellar envelopes and the coupling between gravity and magnetic fields during early stages of star formation.

\end{abstract}

\keywords{Star formation - Interstellar magnetic fields - Circumstellar envelopes - Star forming regions}

\section{Introduction} \label{sec:1} 

A key question for understanding the role of magnetic fields (hereafter,{\em B}-fields) in molecular cloud evolution is how their orientation correlates with the underlying density distribution.
In particular, it is crucial to determine whether {\em B}-fields are preferentially aligned with elongated structures such as filaments, or with density gradients more generally, since these correlations provide insights into the interplay between magnetic support, turbulence, and gravity during cloud collapse \citep[e.g.,][]{ Hennebelle2019magnetic}.

Theories predict that the relative orientation between density structures and {\em B}-fields is not random, but reflects the underlying stability and magnetization of the system.
Linear stability analyses of magnetized, self-gravitating layers \citep{Nagai1998,VanLoo2014} show that the fastest-growing gravitational instability modes depend on the ratio of the scale height $z_{b}$ to the Jeans length $l_{0}$.
When $z_{b} \gg l_{0}$, the layer is compressible and instabilities develop preferentially along field lines, producing filaments that are perpendicular to the {\em B}-field.
Conversely, when $z_{b} \ll l_{0}$, the medium is nearly incompressible and instabilities grow more efficiently with wave vectors perpendicular to the field, yielding filaments aligned with the {\em B}-field.
In the ideal magnetohydrodynamics (MHD) case, {\em B}-fields provide support against gravity through the Lorentz force and channel infalling material toward the equatorial plane, forming flattened pseudo-disks \citep{Galli1993a,Galli1993b}.
As collapse proceeds, the initially uniform field lines are pinched into the familiar hourglass morphology \citep{Nakamura2005}, as observed in both low- and high-mass protostars \citep[e.g.,][]{Girart2006, Girart2009, Stephens2013, Qiu2014, LeGouellec2019}.
Twisted field lines can both drive outflows and remove angular momentum via magnetic braking \citep{Allen2003, Hennebelle2008b}.
This process, however, may lead to the so-called ``magnetic braking catastrophe", in which strong fields prevent the formation of rotationally supported disks \citep{Hennebelle2008a}.
Several mechanisms have been proposed to mitigate this effect, including non-ideal MHD effects that weaken field-matter coupling \citep{Krasnopolsky2002, dapp2010,Krasnopolsky2011}, turbulent diffusion and reconnection \citep{Lazarian1999, SantosLima2012}, and misalignment between the rotation axis and the {\em B}-field \citep{Hennebelle2009, Joos2012, Li2013}.
These scenarios suggest that the magnetization level is a critical parameter that not only regulates disk formation but also influences the overall distribution of material.

Polarized dust emission at (sub)millimeter wavelengths provides a direct probe of {\em B}-fields in dense gas and is the most widely used tracer in star-forming regions. 
In this regime, ``radiative torques" tend to align spinning and elongated dust grains with their long axes perpendicular to the local {\em B}-field \citep{Lazarian2007,Andersson2015}.
This mechanism is effective from cloud core scales ($\sim10^{4}$ au) down to protostellar envelope scales \footnote{Throughout this paper, we use ``protostellar envelope” to refer to material on $\sim10^{2}$--$10^{3}$ au scales around individual protostars} \citep[$\sim10^{2}$--$10^{4}$ au, e.g.,][]{girart1999detection, girart2013dr, Zhang2014, le2020IMS,Pattle2023}.
Recent polarization observational progress has also been made in characterizing the relative orientation between density structures and {\em B}-fields.
\cite{Planck2016XIX} revealed that in the diffuse interstellar medium, elongated column density structures are predominantly aligned with the {\em B}-field in low density regions, while becoming perpendicular in higher density regions, with a transition column density of $\log(N_{\rm H}/{\rm cm^{-2}}) \sim 21.7$. 
Similar statistics are found for low-column-density fibers traced by HI emission \citep{Clark2014}. 
On the scale of molecular clouds and clumps ($0.1$--$1.0$ pc), recent studies \citep[e.g.,][]{Pattle2023} suggest that {\em B}-fields often maintain a high degree of coherence and can even dominate the energy budget, as seen in \cite{Barnes2025}.
\cite{jiao2024structure} reported a similar trend in Orion, where the alignment changes from parallel to perpendicular with increasing column density on both large ($\sim 0.6$ pc) and smaller scales ($\sim 0.04$ pc).
However, these studies are limited to the cloud and core scales ($\gtrsim9000$ au).
It remains an open question whether this density-dependent alignment persists in the dynamic environment of the protostellar envelope, where infall, rotation, and feedback may modify both the density field and magnetic geometry.

The {\em B}-field Orion Protostellar Survey (BOPS; PI: Ian Stephens, project 2019.0.00086S) targets this scale gap, and it mapped polarized dust emission in 57 protostars on 400--3000 au ($\sim$2--15 mpc), revealing three major types of {\em B}-field structures: standard hourglasses (aligned with the outflow), rotated hourglasses (perpendicular to the outflow), and spiral patterns, alongside complex configurations \citep{Huang2024}.
These morphologies appear to correlate with magnetization levels \citep{huang2025a, huang2025b}, providing an opportunity to investigate how {\em B}-fields shape density structures when the magnetic support varies relative to gravity.
This paper is organized as follows: Section~\ref{sec:2} describes the BOPS observations, Section~\ref{sec:3} presents the analysis methods and results, Section~\ref{sec:4} discusses the implications for magnetic regulation of protostellar collapse, and Section~\ref{sec:5} summarizes the main conclusions.

\begin{figure*}
\centering 
\includegraphics[width=0.68\linewidth]{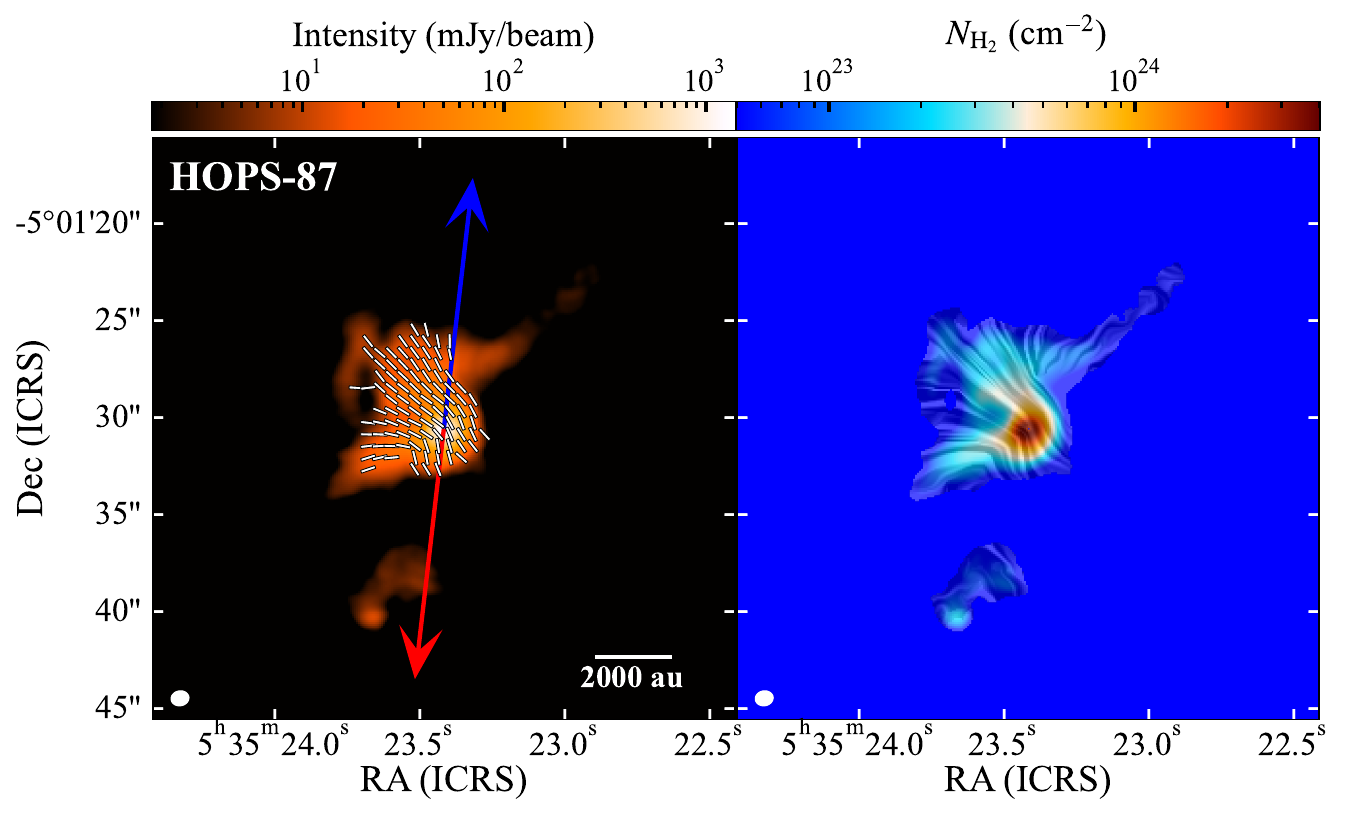}\\
\includegraphics[width=0.68\linewidth]{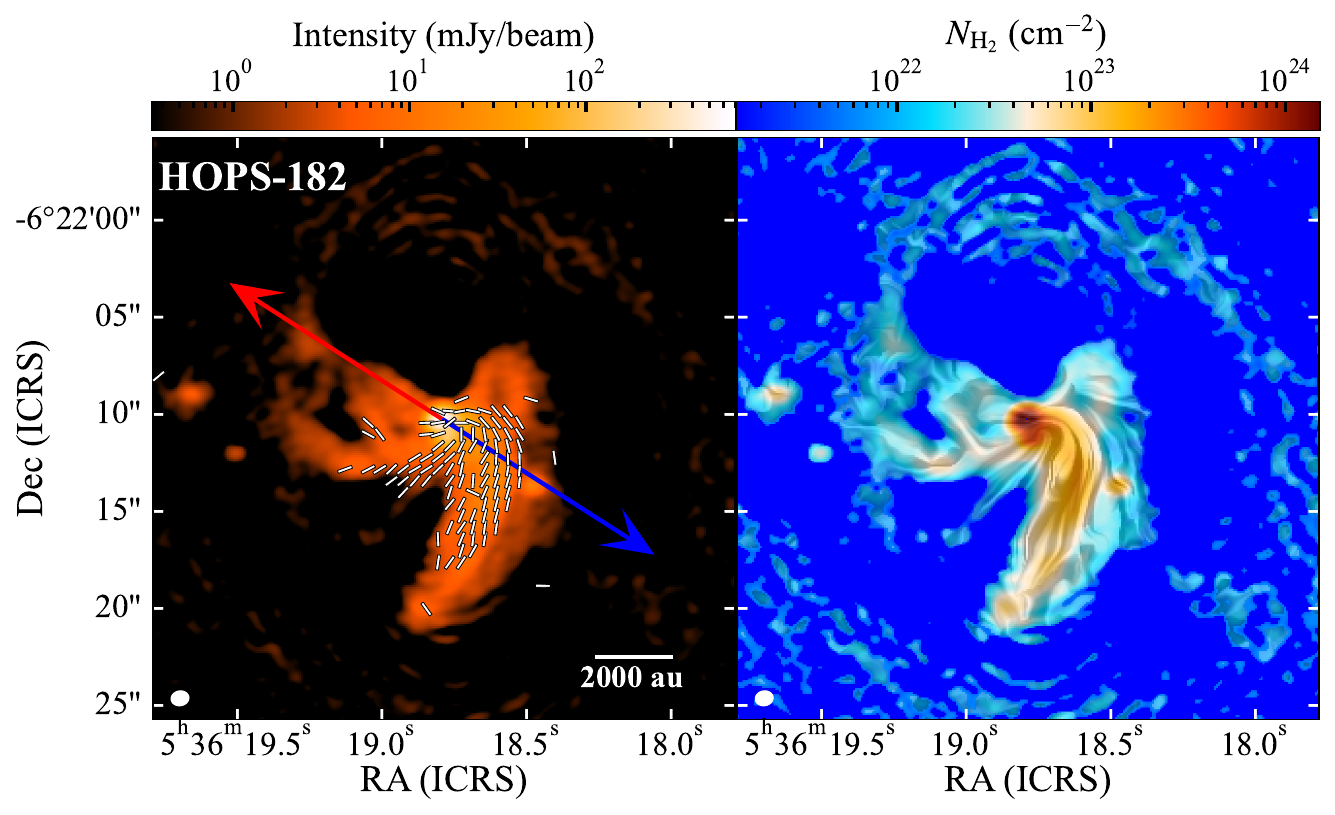}\\
\includegraphics[width=0.68\linewidth]{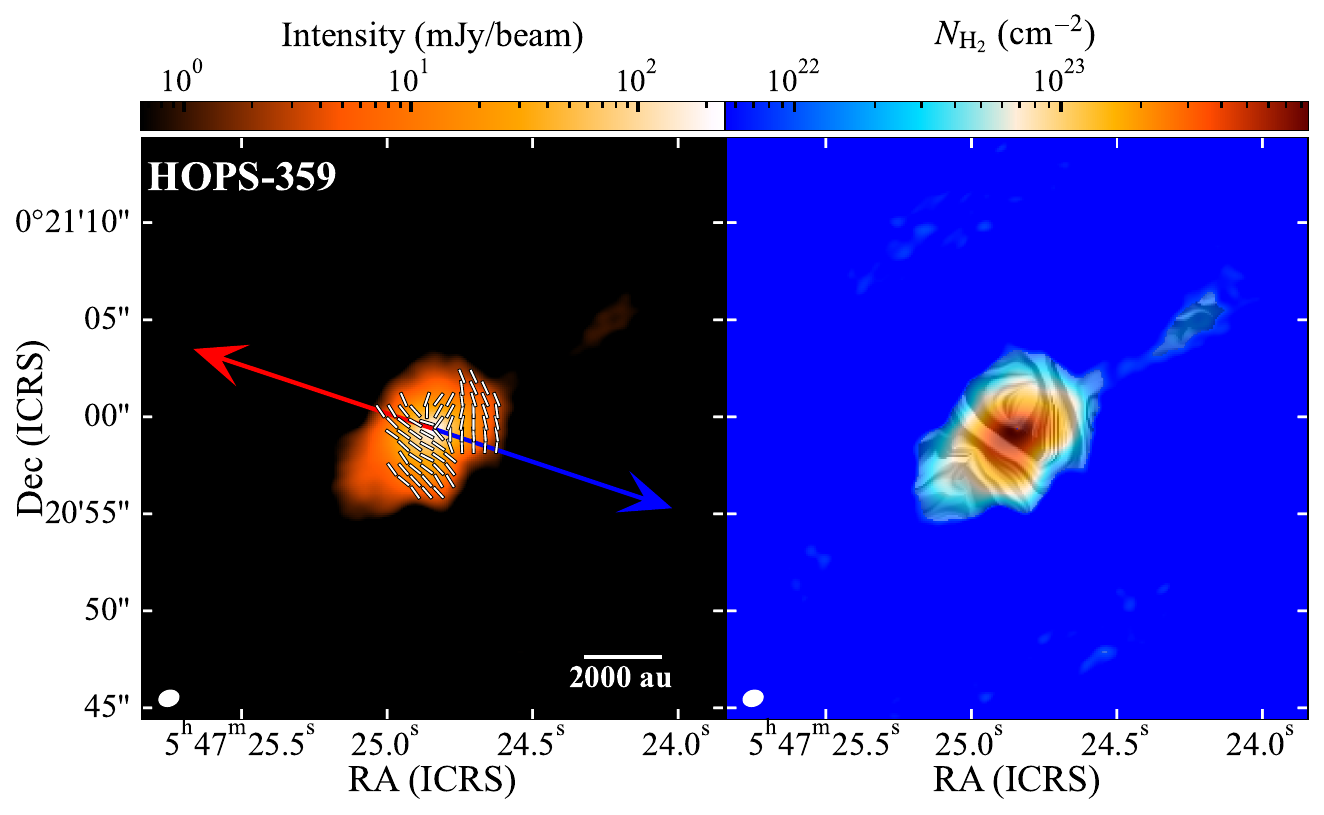}\\
\text{\textbf{{Figure 1.}} ALMA BOPS 870 $\mu$m observations (Continued.)}
\end{figure*}

\begin{figure*}
\centering 
\includegraphics[width=0.68\linewidth]{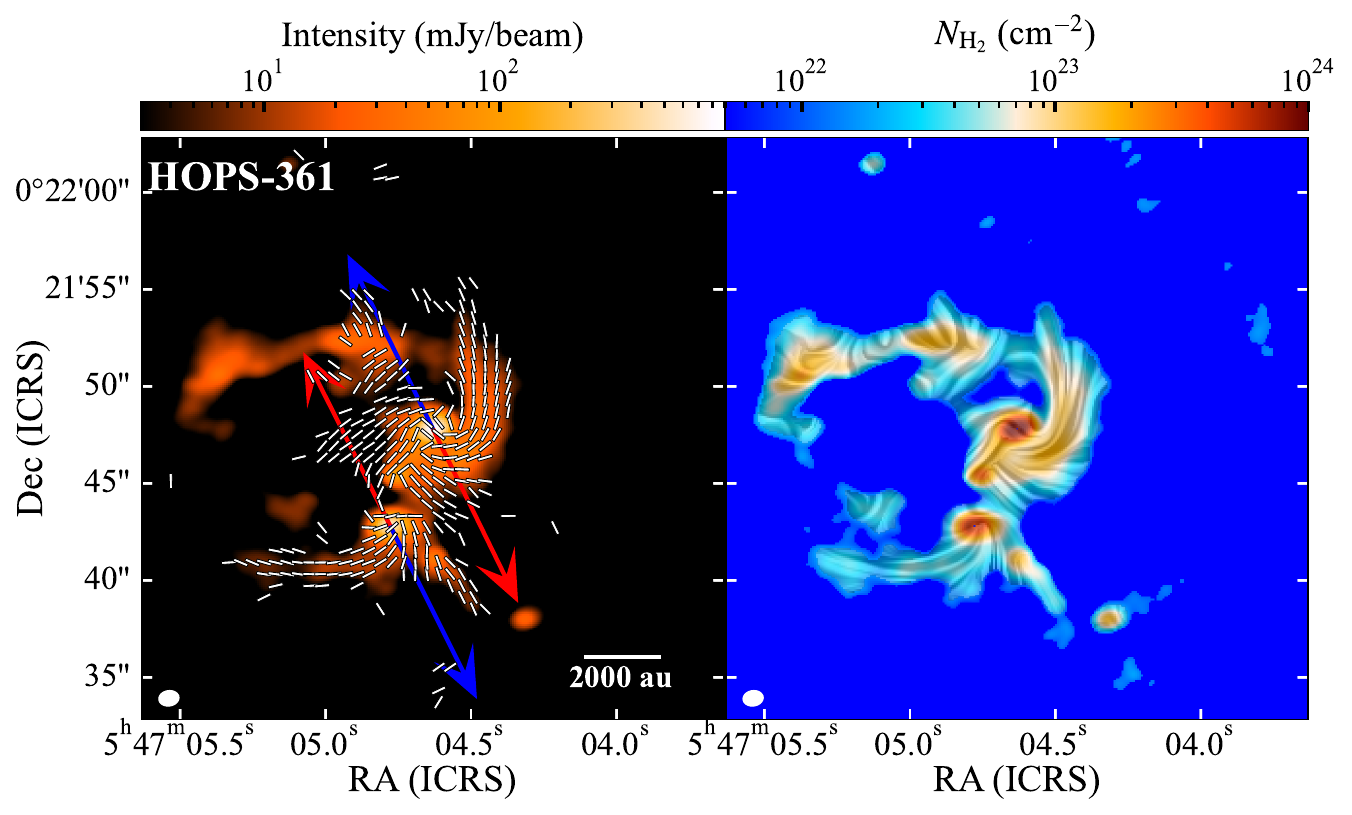}\\
\includegraphics[width=0.68\linewidth]{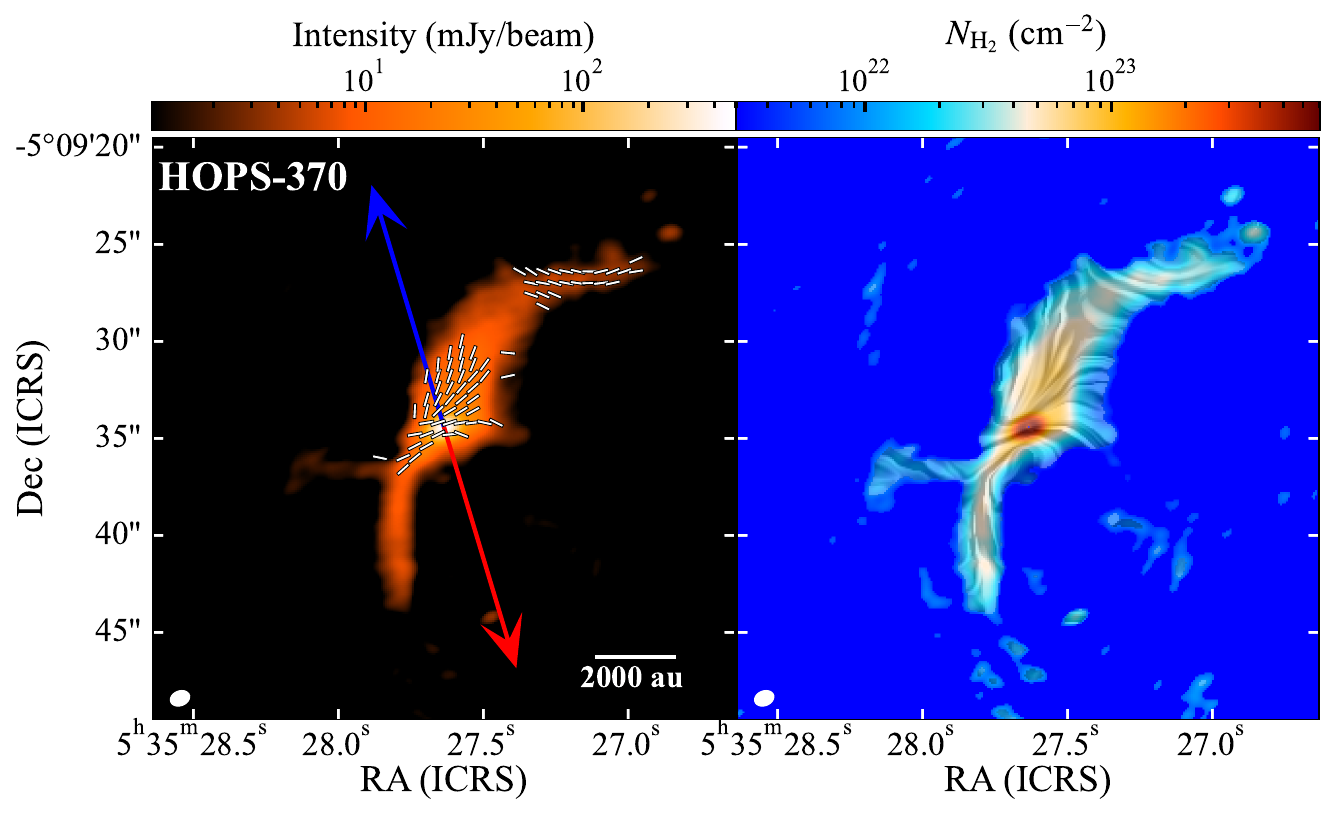}\\
\includegraphics[width=0.68\linewidth]{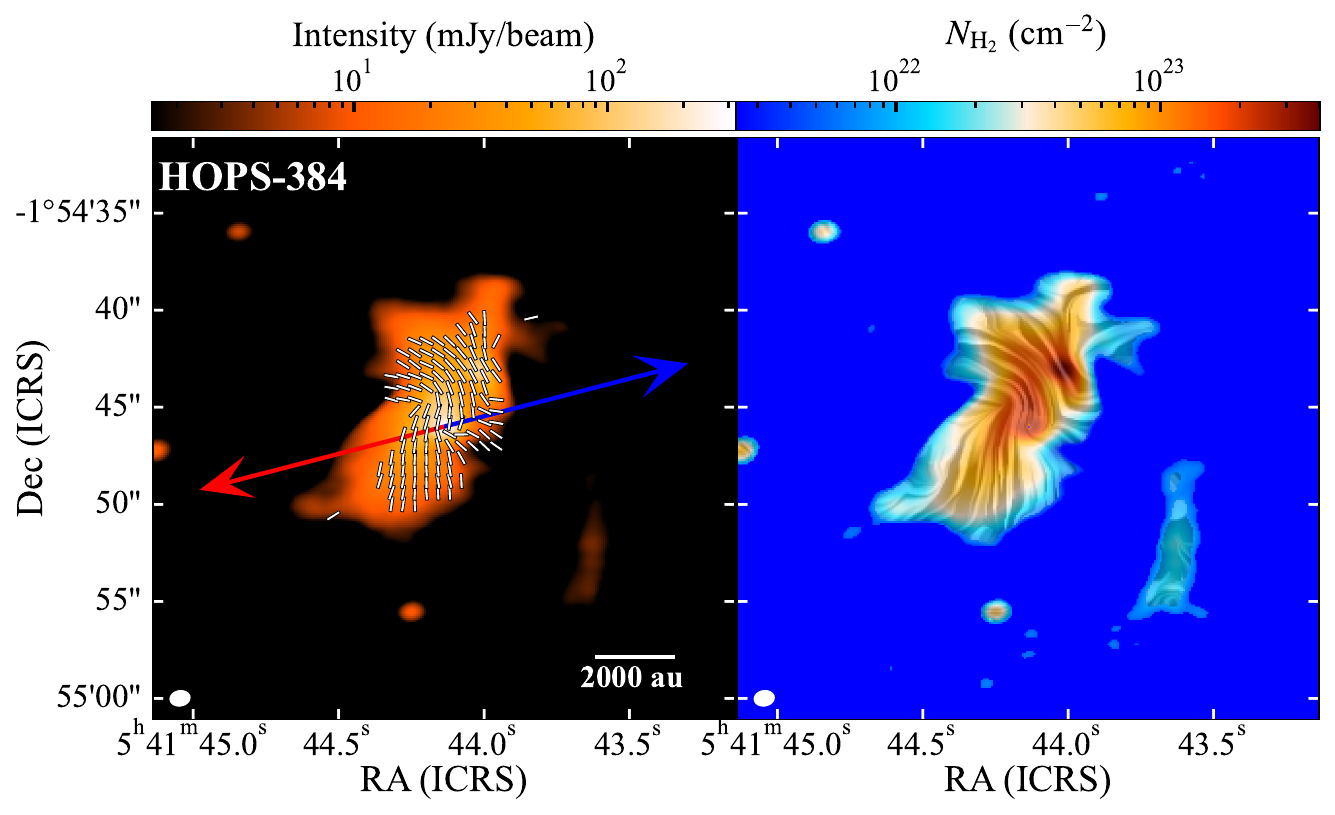}\\
\text{\textbf{{Figure 1.}} ALMA BOPS 870 $\mu$m observations (Continued.)}
\end{figure*}

\begin{figure*}
\centering 
\includegraphics[width=0.68\linewidth]{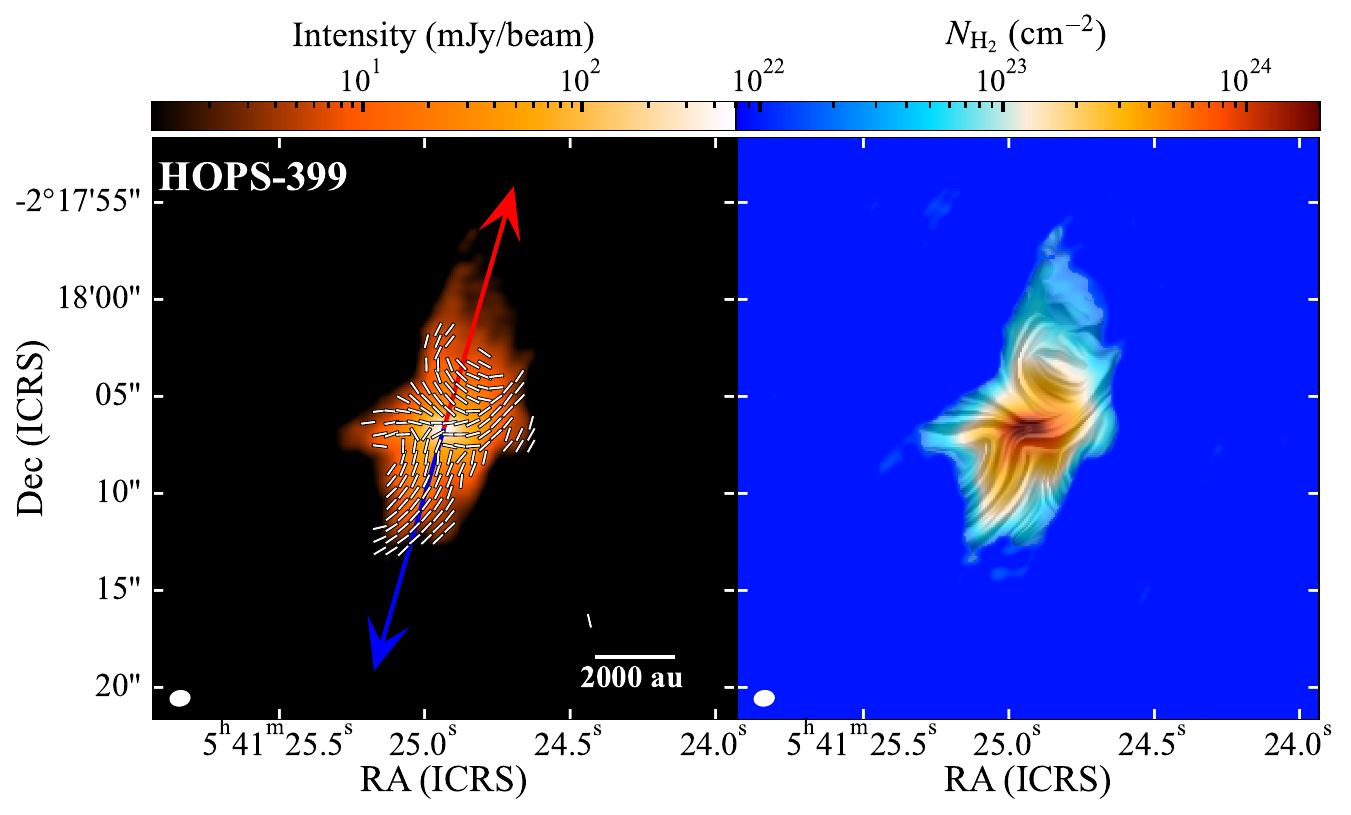}\\
\includegraphics[width=0.68\linewidth]{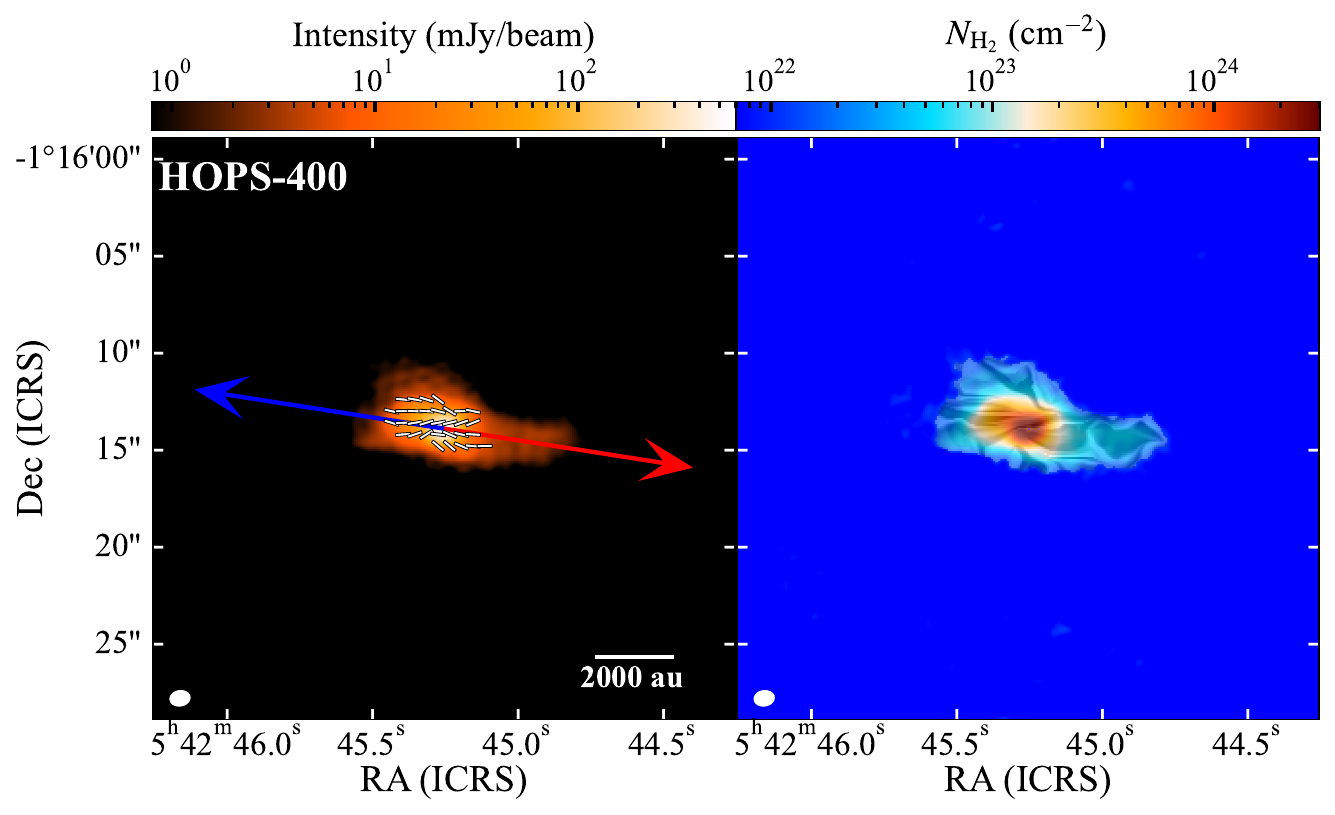}\\
\caption{ALMA BOPS 870 $\mu$m observations.
{\em Left panel}: The dust continuum emission (Stokes $I$) is shown in color scales, overlaid with the {\em B}-field segments in white and the redshifted and blueshifted outflow directions in red and blue, respectively. 
{\em Right panel}: Column density maps derived from the 870 $\mu$m continuum emission in color scale, overlaid with an LIC texture of the inferred plane-of-sky {\em B}-field orientation computed from unit polarization-angle vectors, i.e., the LIC texture encodes direction \citep{cabral1993special}.}
\label{fig:obs} 
\end{figure*}

\section{Observations} \label{sec:2}

The BOPS observations were carried out between November and December 2019 using the ALMA 12 m array in the compact configurations C43-1 and C43-2 (and an intermediate execution combining both), which provided projected baselines of $\sim$14--312 m. The observations were obtained in Band~7 ($\sim$345\,GHz; 870~$\mu$m). 
The resulting typical synthesized beam is $\sim 0.8^{\prime\prime}\times0.6^{\prime\prime}$, while the largest angular scale (LAS) reaches $\sim 8^{\prime\prime}$. 
The data reduction followed the standard ALMA pipeline in CASA, and imaging was performed with the \texttt{tclean} task using Briggs weighting (robust = 0.5). We applied three successive rounds of phase-only self-calibration to improve image quality for each source. 
The final Stokes $I$, $Q$, and $U$ continuum maps were produced independently using line-free, self-calibrated data from four spectral windows. The peak S/N in the final Stokes $I$ maps ranges from $\sim$750 to $\sim$4200, and the typical rms noise levels are $\sim$0.1\,mJy\,beam$^{-1}$ in Stokes $I$ and $\sim$0.07\,mJy\,beam$^{-1}$ in Stokes $Q$ and $U$.
From the polarization maps we derive the polarized intensity $P=\sqrt{Q^{2}+U^{2}}$ and the polarization position angle $\theta_{p}=0.5\arctan(U/Q)$, where $I$, $Q$, and $U$ are the Stokes parameters.
By definition, $P$ is always positive, even though $Q$ and $U$ can take positive and negative values.
This introduces a positive bias in low-signal-to-noise measurements, which we correct by debiasing the polarized intensity following the method described by \cite{hull2015debias}.

We focus on eight protostars, i.e., HOPS-87, HOPS-182, HOPS-359, HOPS-361, HOPS-370, HOPS-384, HOPS-399, and HOPS-400, that not only have bolometric luminosities reported in the literature for column density estimates \citep{furlan2016herschel}, but also cover the diverse {\em B}-field morphologies and exhibit sufficient polarization detections (from $\sim15$ to $\sim150$ independent beams with $P\geq3\sigma$) to allow a robust Histogram of Relative Orientations \citep[HRO;][]{Soler2013} analysis. 
Among them, two (HOPS-87 and HOPS-400) display standard hourglass structures, one (HOPS-359) shows a partially hourglass configuration, one (HOPS-370) a rotated hourglass, three (HOPS-182, HOPS-361, and HOPS-384) spiral morphologies, and one (HOPS-399) a complex pattern \citep{Huang2024}.
The left panels of Figure~\ref{fig:obs} show the 870 $\mu$m dust emission in color scale for these sources, overlaid with {\em B}-field segments ($P\geq3\sigma$ detections) and outflow directions. 

\begin{table}
\centering
\caption{Source properties.
Columns (2)-(6) present the {\em B}-field morphologies, bolometric luminosity ($L_{\rm bol}$), overall trend of histogram shape parameter $\xi$, alignment preference between {\em B}-field and density structures, and overall trend of {\em B}-field structure function, respectively.
We classify sources as ``small dispersion" when the SF amplitude remains below the random-field expectation at all lags (i.e. $<52^{\circ}$), and as ``large-dispersion" otherwise.
The type of {\em B}-field is obtained from \cite{Huang2024}, 
$L_{\rm bol}$ is adopted from \cite{furlan2016herschel}. }
\begin{tabular}{ccccccccccc}
\hline
\hline
Name & ~ & Type of {\em B}-field & ~ & $L_{\rm bol}~(L_{\odot})$ & ~ & $\xi$ & ~ & Dominant Alignment & ~ & SF \\
\hline
HOPS-87 & ~~ & std-hourglass & ~~ & 36.49 & ~~ & predominantly negative & ~~ & perpendicular & ~~ & small dispersion  \\
HOPS-182 & ~ & spiral & ~ & 71.12 & ~ & predominantly positive & ~ & parallel & ~ & large dispersion  \\
HOPS-359 & ~ & part-hourglass & ~ & 10.00 & ~ & scatter near 0 & ~ & no clear trend & ~ & small dispersion  \\
HOPS-361 & ~ & spiral & ~ & 478.99 & ~ & predominantly positive & ~ & parallel & ~ & large dispersion  \\

HOPS-370 & ~~ & rot-hourglass & ~~ & 360.86 & ~~ & predominantly positive & ~~ & parallel & ~~ & large dispersion  \\
HOPS-384 & ~ & spiral & ~ & 1477.99 & ~ & negative to positive & ~ & Perp.-to-Para. & ~ & large dispersion  \\
HOPS-399 & ~ & complex & ~ & 6.34 & ~ & predominantly positive & ~ & parallel & ~ & large dispersion  \\
HOPS-400 & ~ & std-hourglass & ~ & 2.94 & ~ & predominantly negative & ~ & perpendicular & ~ & small dispersion  \\
\hline
\end{tabular}
\label{tab:source}
\end{table}

\section{Results and analysis} 
\label{sec:3}

In this section, we describe the procedures used to compare the density structure with the {\em B}-field orientation, including the derivation of column density maps, the application of the HRO method and the calculation of the {\em B}-field structure function.

\subsection{Column density maps} \label{subsec:3.1}
\begin{figure*}
\centering 
\includegraphics[width=0.48\linewidth]{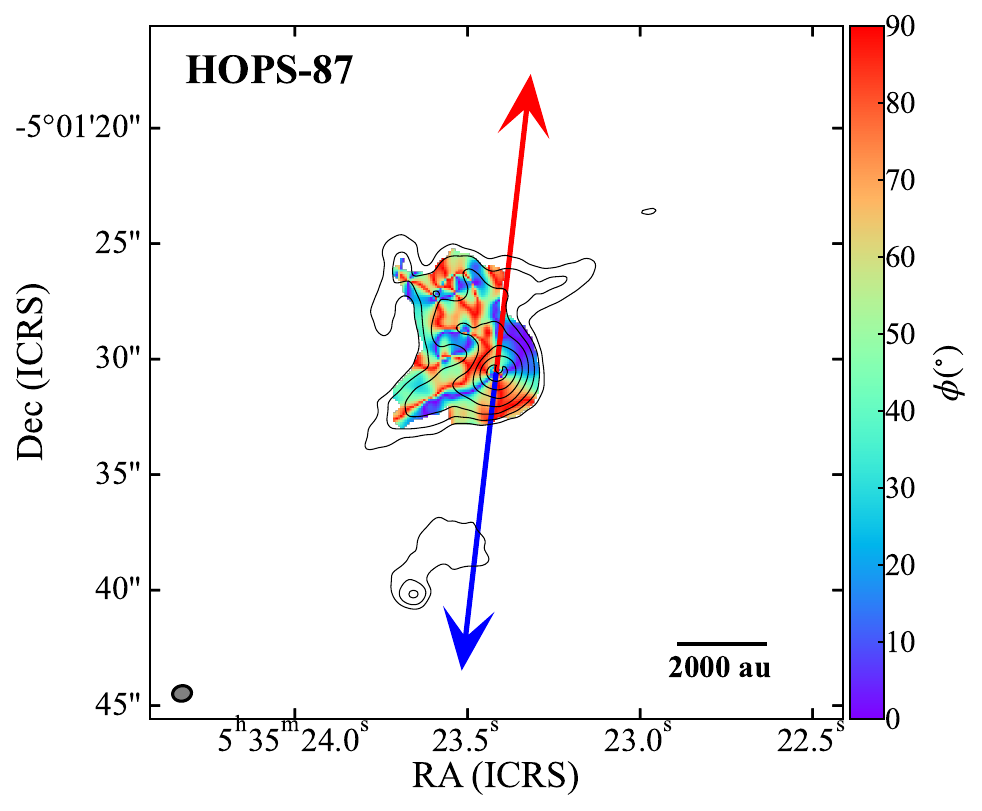}~~~~~
\includegraphics[width=0.48\linewidth]{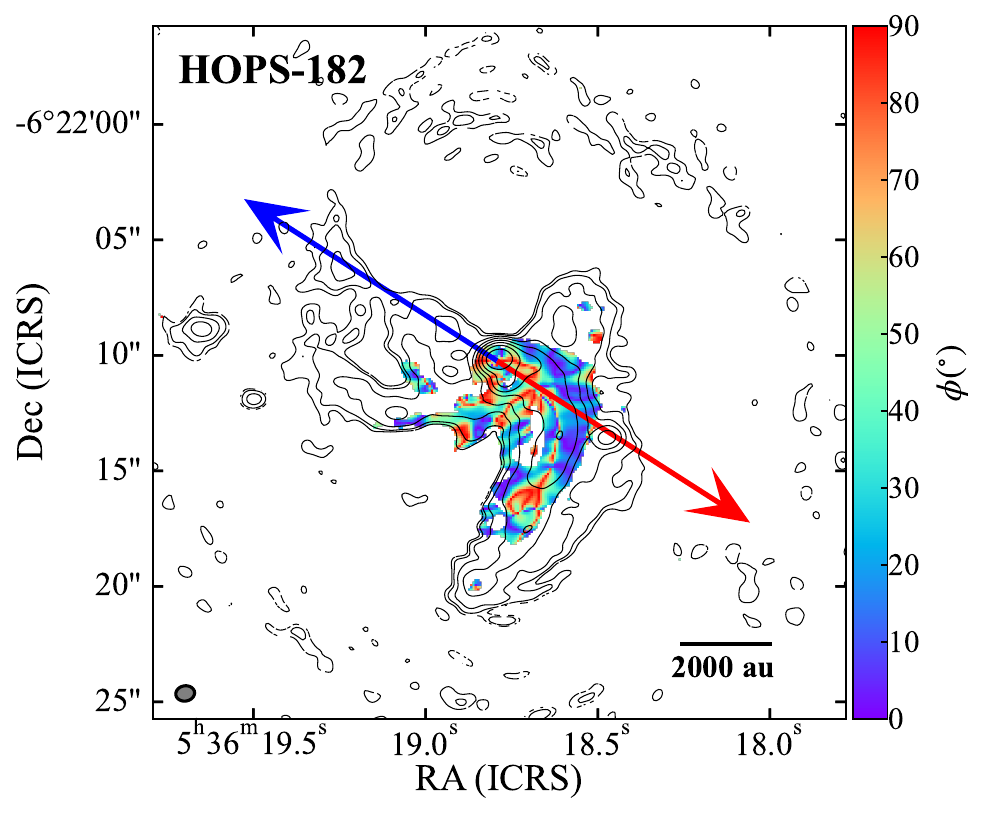}
\includegraphics[width=0.48\linewidth]{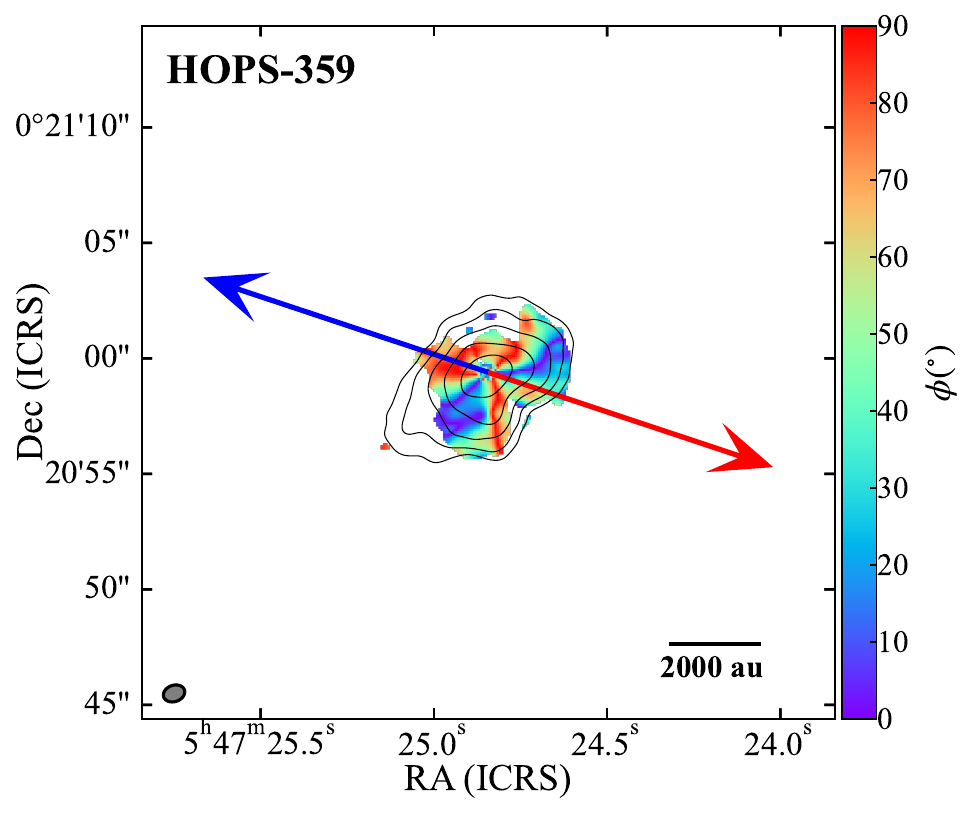}~~~~~
\includegraphics[width=0.48\linewidth]{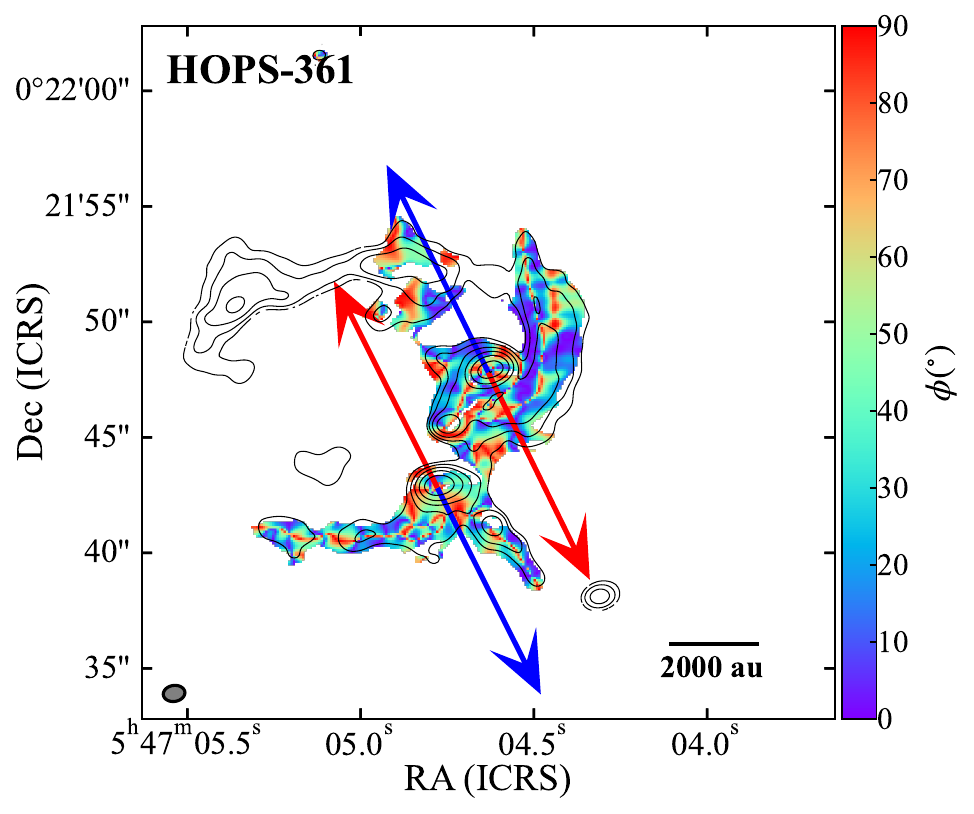}
\includegraphics[width=0.48\linewidth]{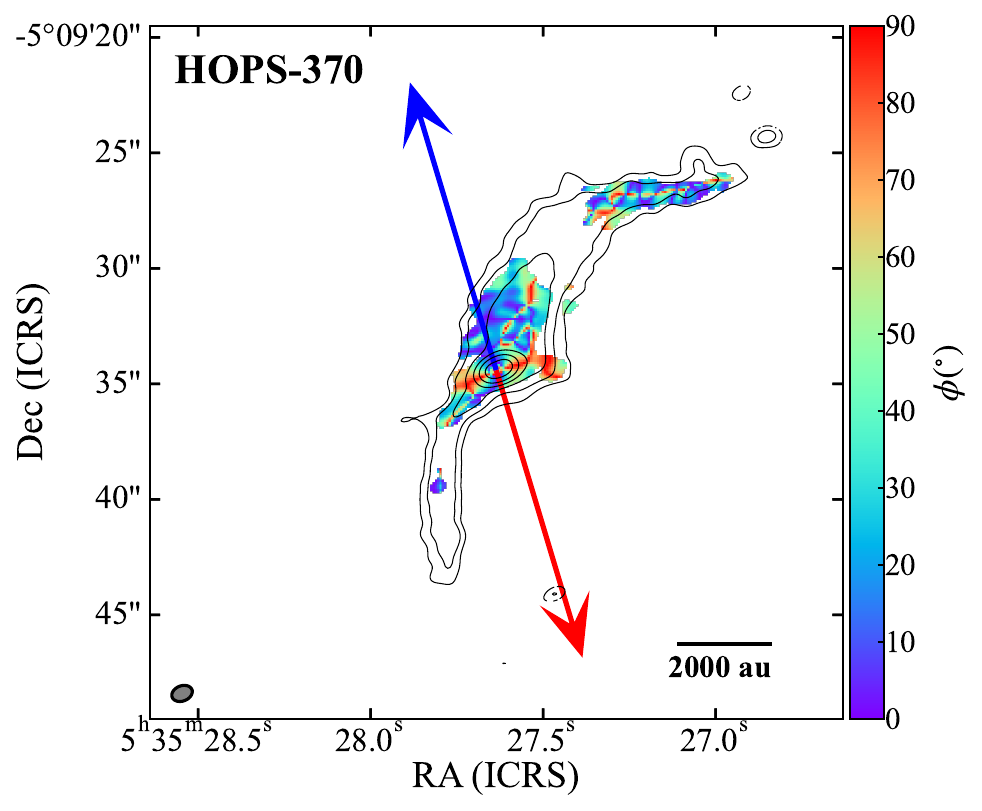}~~~~~
\includegraphics[width=0.48\linewidth]{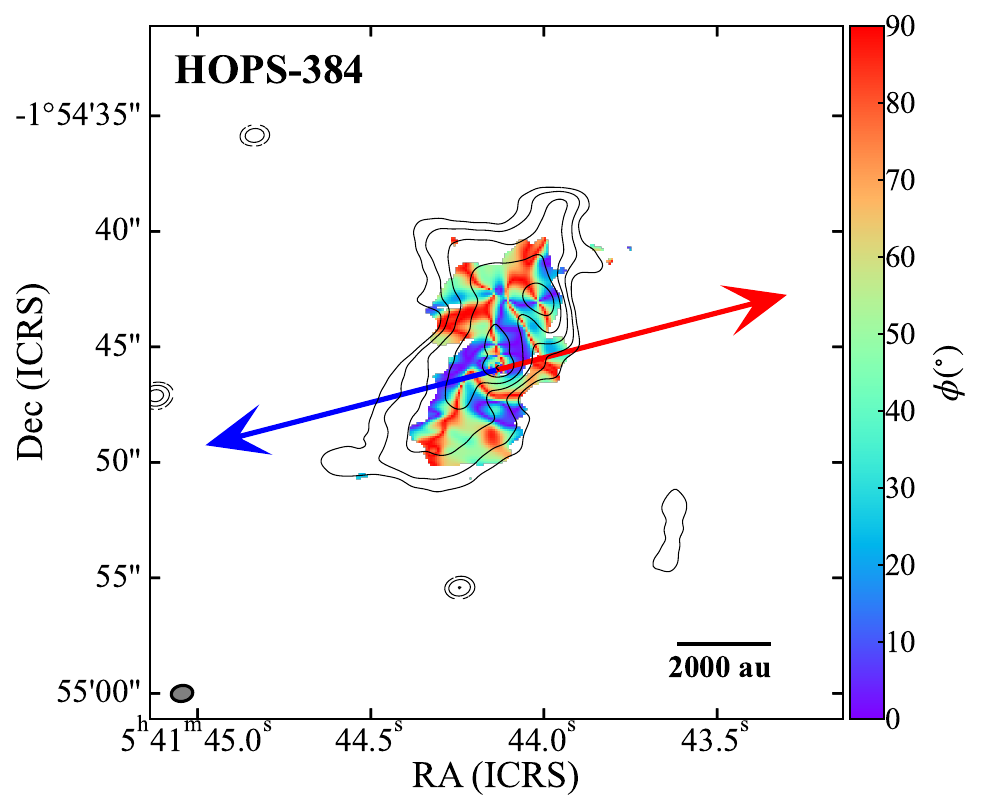}
\text{\textbf{{Figure 2.}} The relative alignment $\phi$ between the column density contours and the {\em B}-field orientation (Continued.)}
\end{figure*}


\begin{figure*}
\centering 
\includegraphics[width=0.48\linewidth]{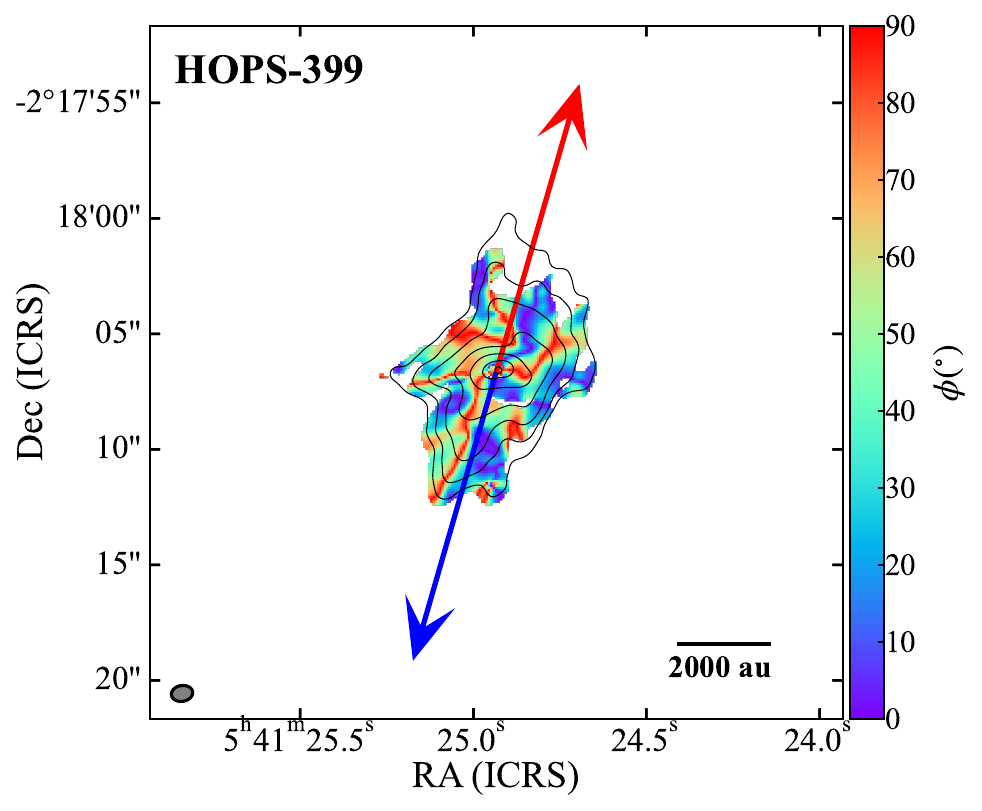}~~~~~
\includegraphics[width=0.48\linewidth]{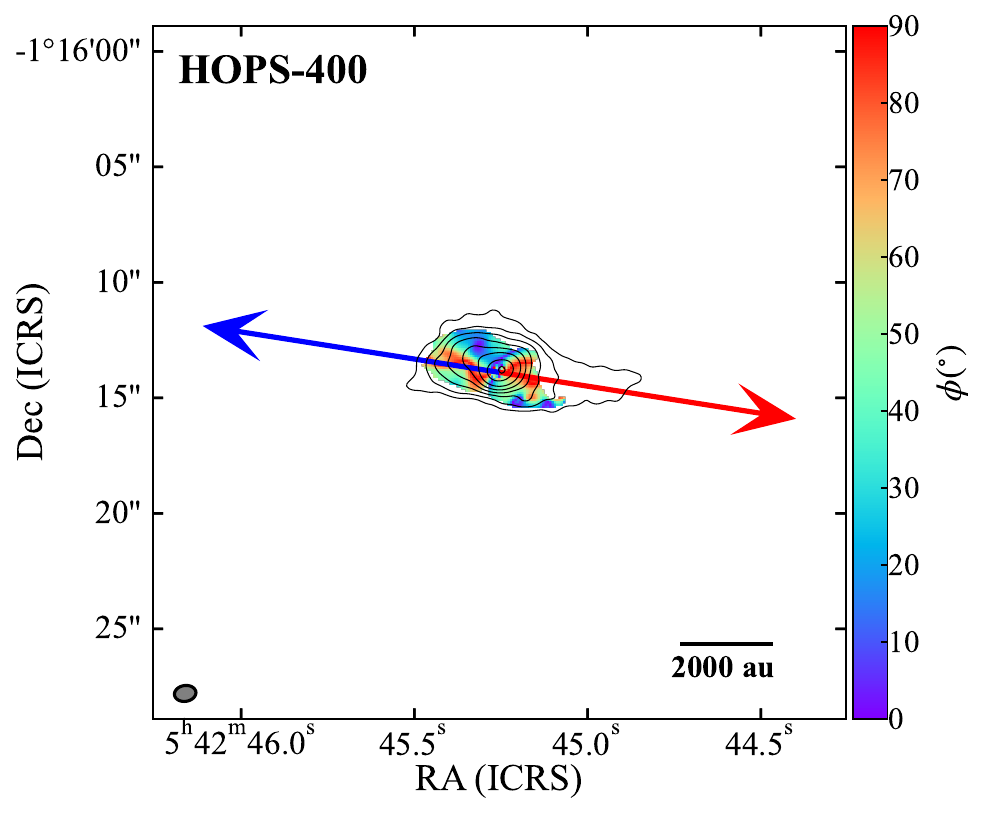}
\caption{The relative alignment $\phi$ between the column density contours and the {\em B}-field orientation in color scales, overlaid with the column density contours.
Contours in each panel start at the 5$\sigma$ ($\sigma$ is the noise level in column density) and increase by factors of 2 (i.e., 1, 2, 4, 8, 16, 32, 64, 128 $\times~ 5\sigma$), with the noise level of $9.9\times10^{21}~{\rm cm^{-2}}$ for HOPS-87, $1.3\times10^{21}~{\rm cm^{-2}}$ for HOPS-182, $5.4\times10^{21}~{\rm cm^{-2}}$ for HOPS-359, 
$4.7\times10^{21}~{\rm cm^{-2}}$ for HOPS-361, $2.6\times10^{21}~{\rm cm^{-2}}$ for HOPS-370, $2.5\times10^{21}~{\rm cm^{-2}}$ for HOPS-384, $9.7\times10^{21}~{\rm cm^{-2}}$ for HOPS-399, and $7.0\times10^{21}~{\rm cm^{-2}}$ for HOPS-400.}    
\label{fig:angle_difference}  
\end{figure*}

Assuming the dust emission is optically thin on envelope scales ($\sim 10^{3}$ au), the column density is computed from the 870 $\mu$m dust continuum using
\begin{equation}\label{eq1}
N_{\rm H_{2}}=\eta\frac{S_{\nu}d^{2}}{\mu_{\rm H_{2}}m_{\rm H}\kappa_{\nu}B_{\nu}(T)},
\end{equation}
where $S_{\nu}$ is the flux density at a frequency $\nu$ = 345 GHz (i.e., a wavelength of 870~$\mu$m), $d$ is the distance, $\mu_{\rm H_{2}}=2.8$ is the mean molecular weight per hydrogen molecule \citep{kauffmann2008mambo}, $m_{\rm H}$ is the mass of the hydrogen atom, $\kappa_{\nu}$ is the dust opacity, $\eta=100$ is the gas-to-dust mass ratio, and $B_{\nu}(T)$ is the Planck function at dust temperature $T$.
We adopt $\kappa_{\nu}\approx1.84~{\rm cm^{2}~g^{-1}}$, appropriate for grains with ice mantles at densities of $\sim10^{6}~{\rm cm^{3}}$ \citep{ossenkopf1994dust}.

The dust temperature profile for very young Orion protostars on envelope scales is expressed as \citep{huang2025a}
\begin{equation}\label{eq2}
T = T_{0}\biggl(\frac{L_{\rm bol}}{L_{\odot}}\biggr)^{0.25}\biggl(\frac{r}{50~{\rm au}}\biggr)^{-0.40} \, ,
\end{equation}
where $T_{0}=43$ K corresponds to the average temperature of a $\sim1~L_{\odot}$ protostar at $r\sim50$ au \citep{whitney2003radiative, tobin2013resolved, tobin2020vla},  $r$ is the distance from the protostar, and $L_{\rm bol}$ is the bolometric luminosity taken from \cite{furlan2016herschel} (column (3) of Table~\ref{tab:source}).
Since C$^{17}$O emission is clearly detected in our sample, the dust temperature cannot fall below 20 K, as C$^{17}$O would otherwise freeze onto dust grains and remain undetectable \citep[e.g.,][]{jorgensen2015molecule}. 
In cases where Eq.~\ref{eq2} yields temperatures below this value, we adopt 20 K as a lower limit.
The right panels of Figure~\ref{fig:obs} present the resulting column density maps in color scale, overlaid with the {\em B}-field morphology generated using the linear integral convolution (LIC) technique \citep{cabral1993special}.

\subsection{Calculation of relative orientations
}\label{subsec:3.2}

\begin{figure*}
\centering 
\includegraphics[width=0.32\linewidth]{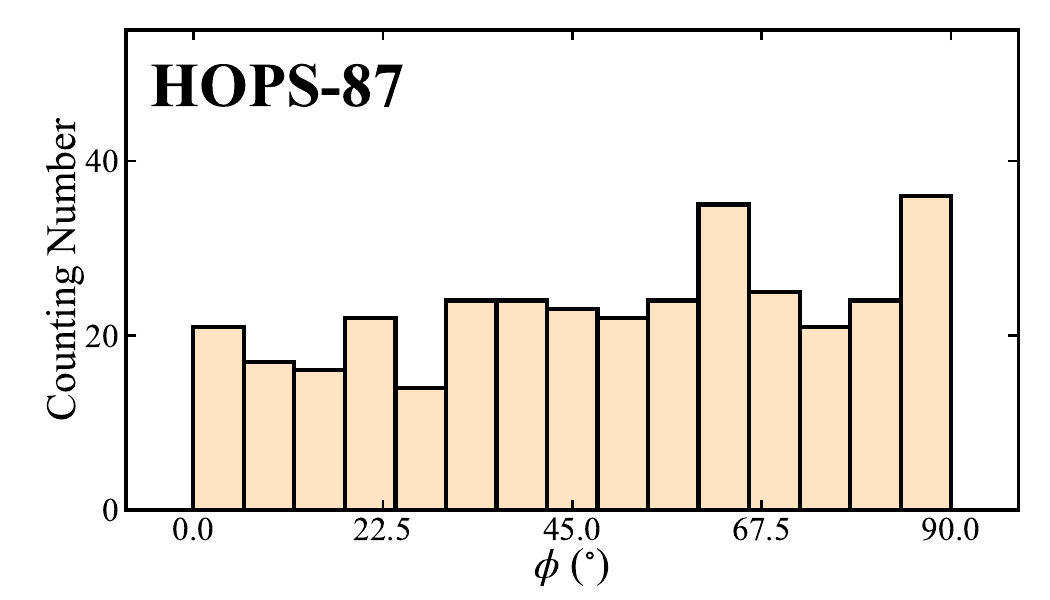}~
\includegraphics[width=0.32\linewidth]{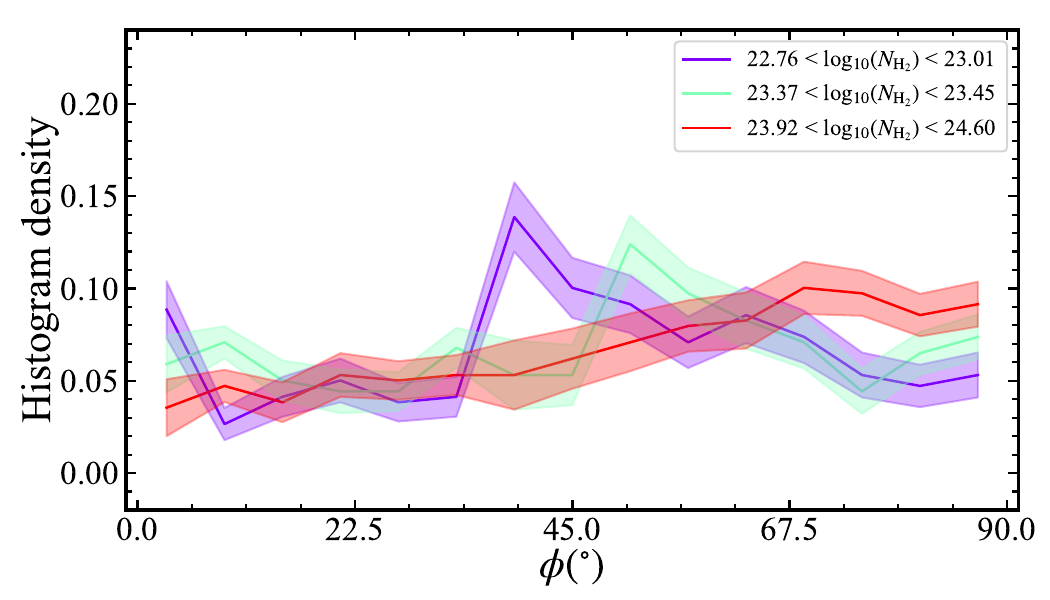}~
\includegraphics[width=0.32\linewidth]{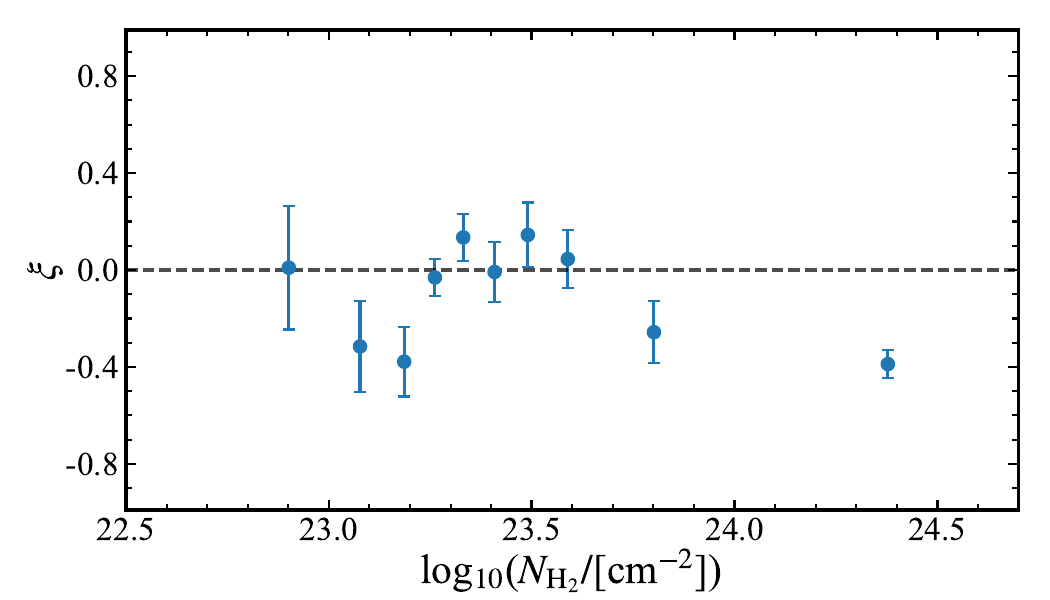}\\
\vspace{0.21cm}
\includegraphics[width=0.32\linewidth]{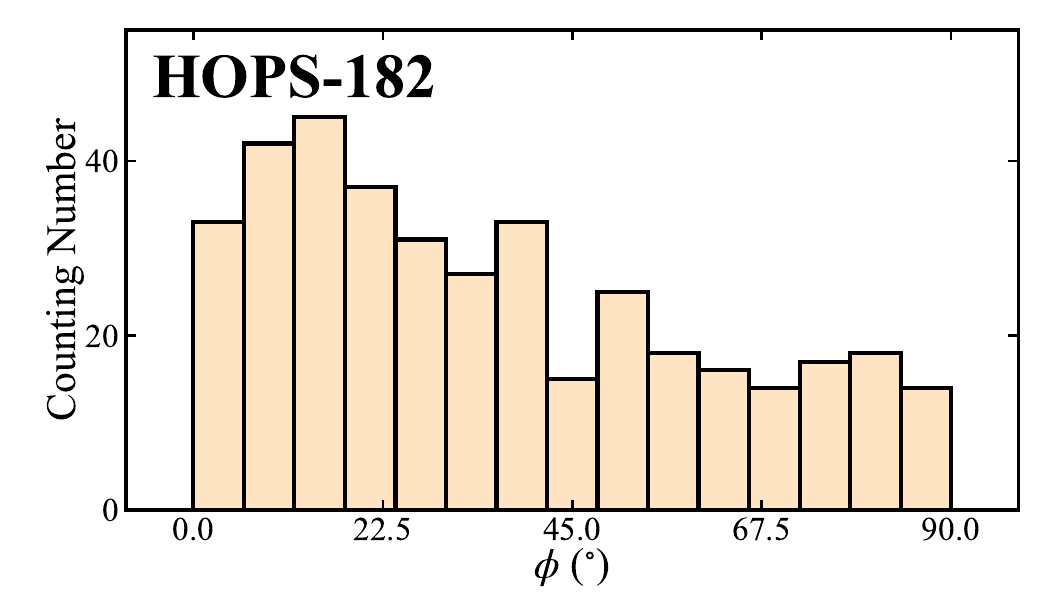}~
\includegraphics[width=0.32\linewidth]{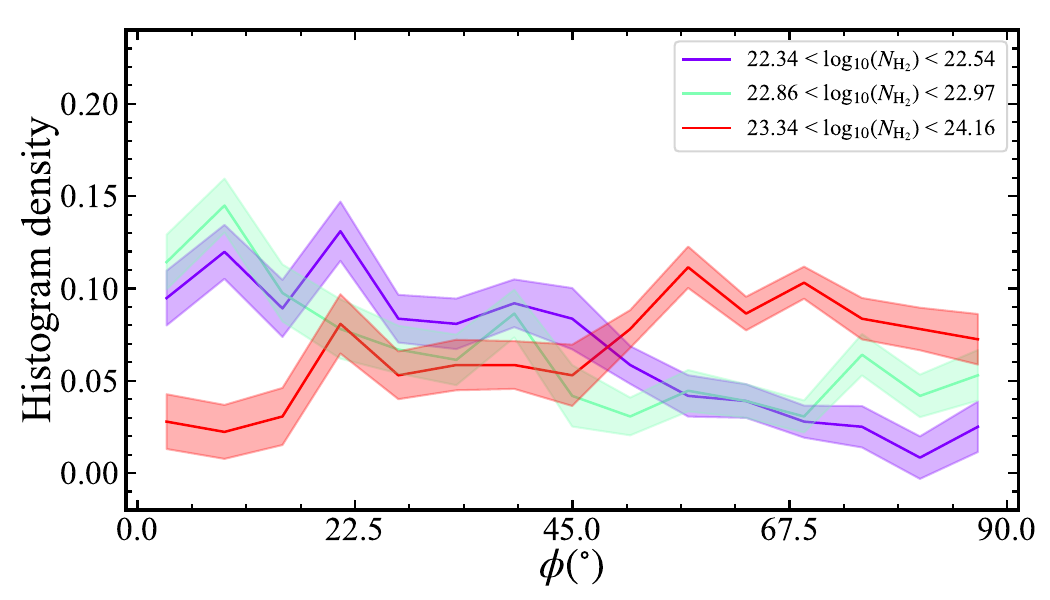}~
\includegraphics[width=0.32\linewidth]{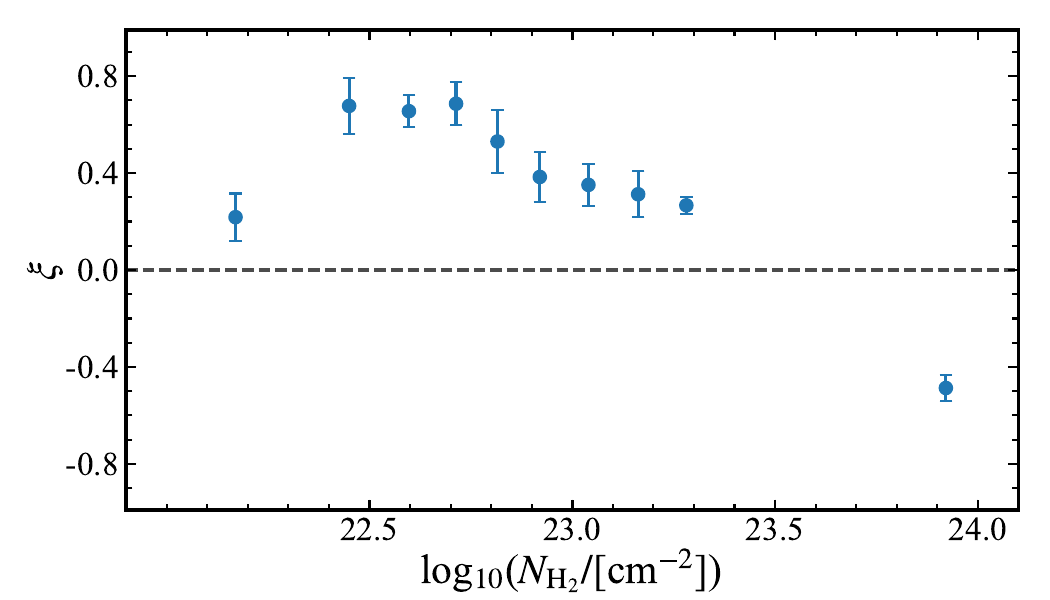}\\
\vspace{0.21cm}
\includegraphics[width=0.32\linewidth]{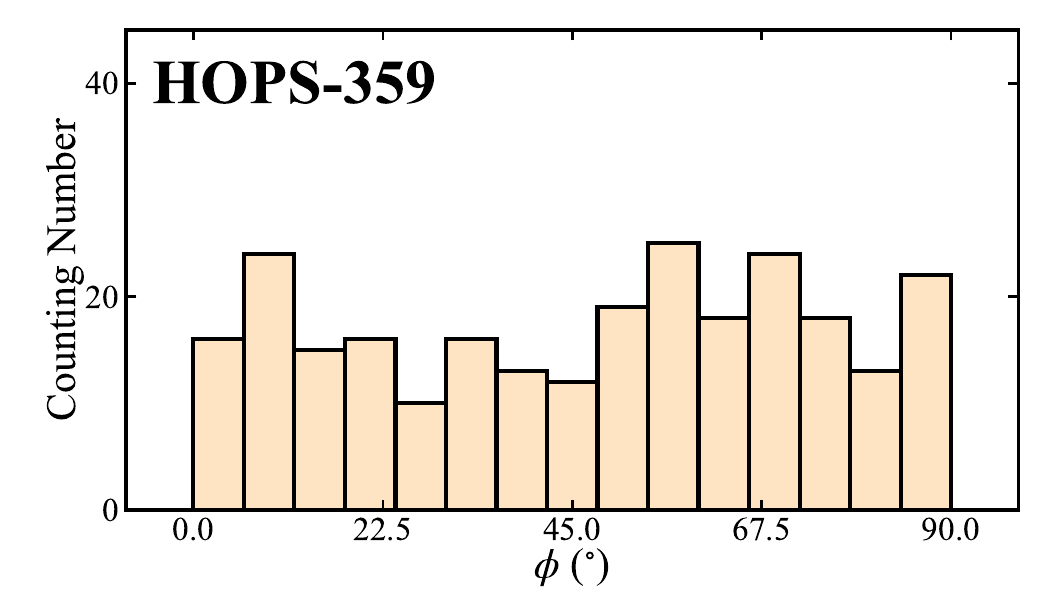}~
\includegraphics[width=0.32\linewidth]{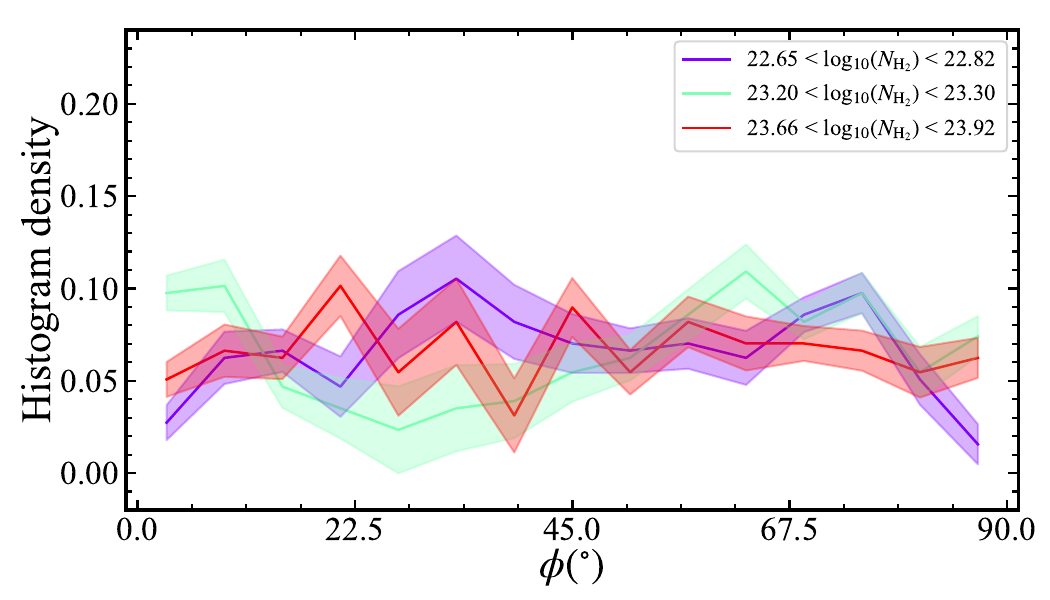}~
\includegraphics[width=0.32\linewidth]{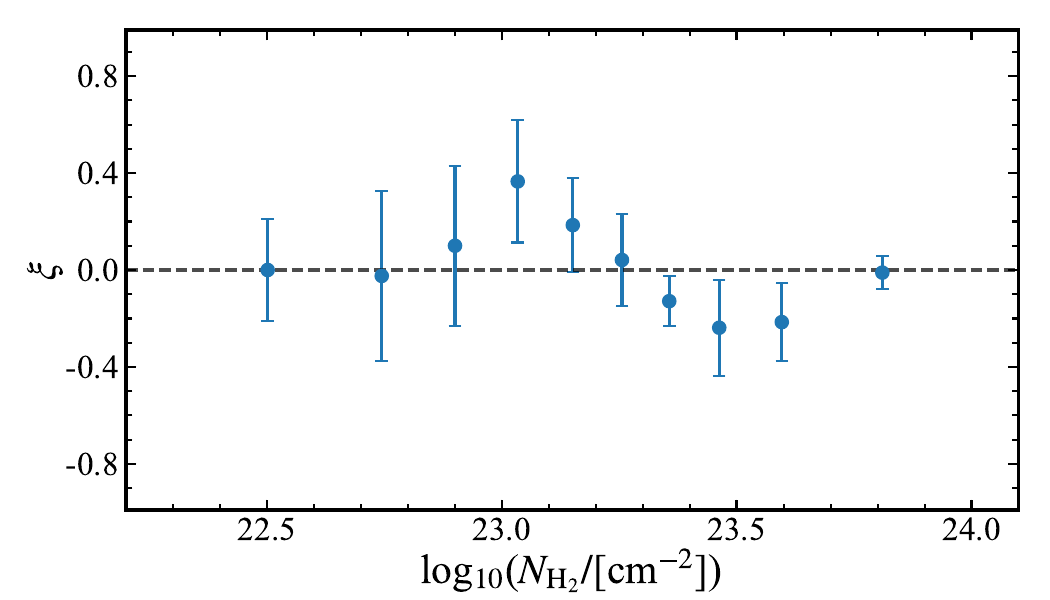}\\
\vspace{0.21cm}
\includegraphics[width=0.32\linewidth]{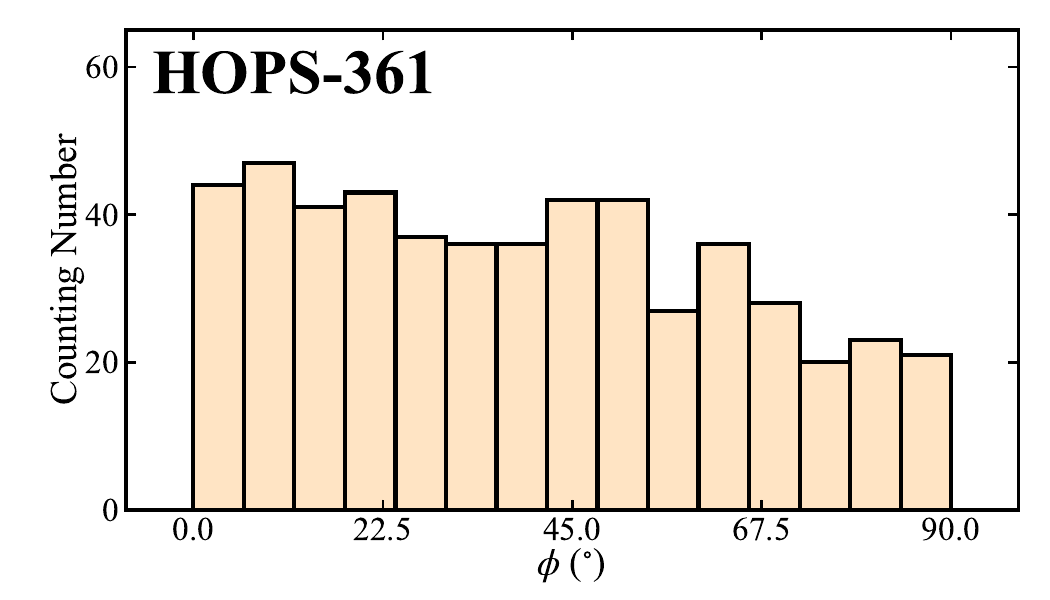}~
\includegraphics[width=0.32\linewidth]{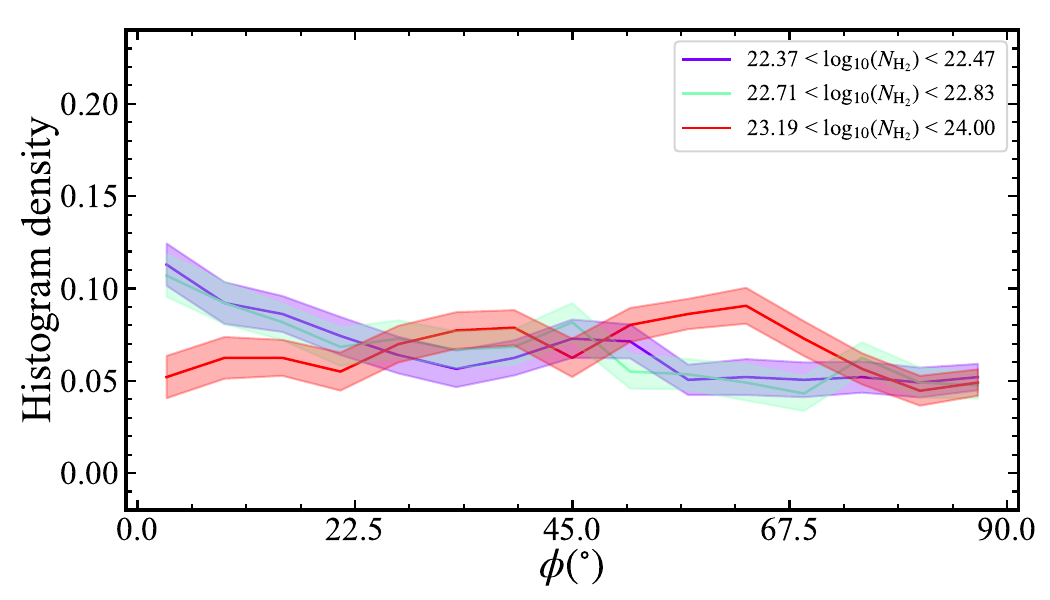}~
\includegraphics[width=0.32\linewidth]{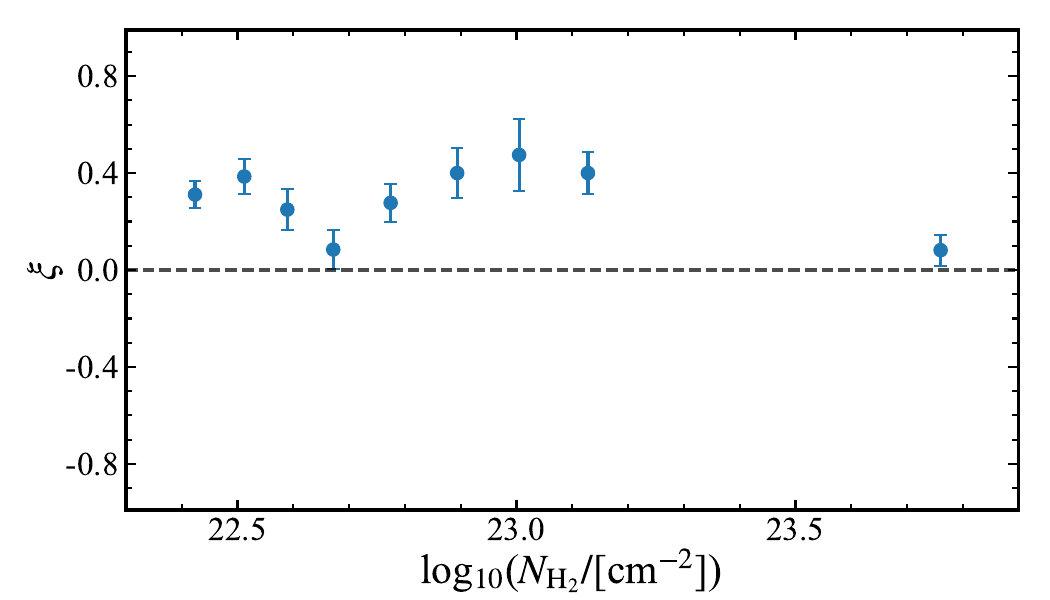}\\
\vspace{0.21cm}
\includegraphics[width=0.32\linewidth]{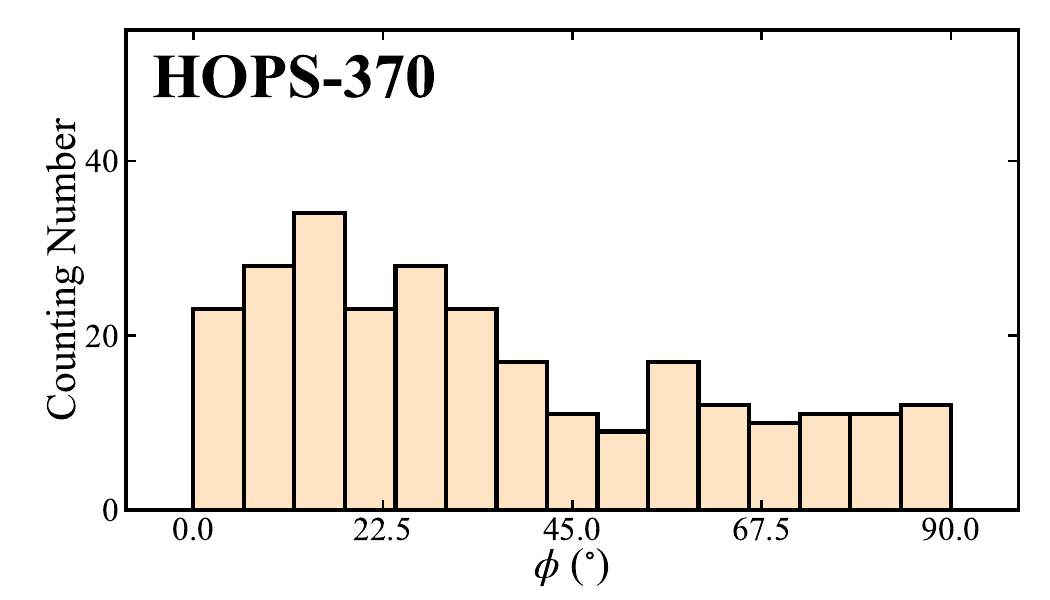}~
\includegraphics[width=0.32\linewidth]{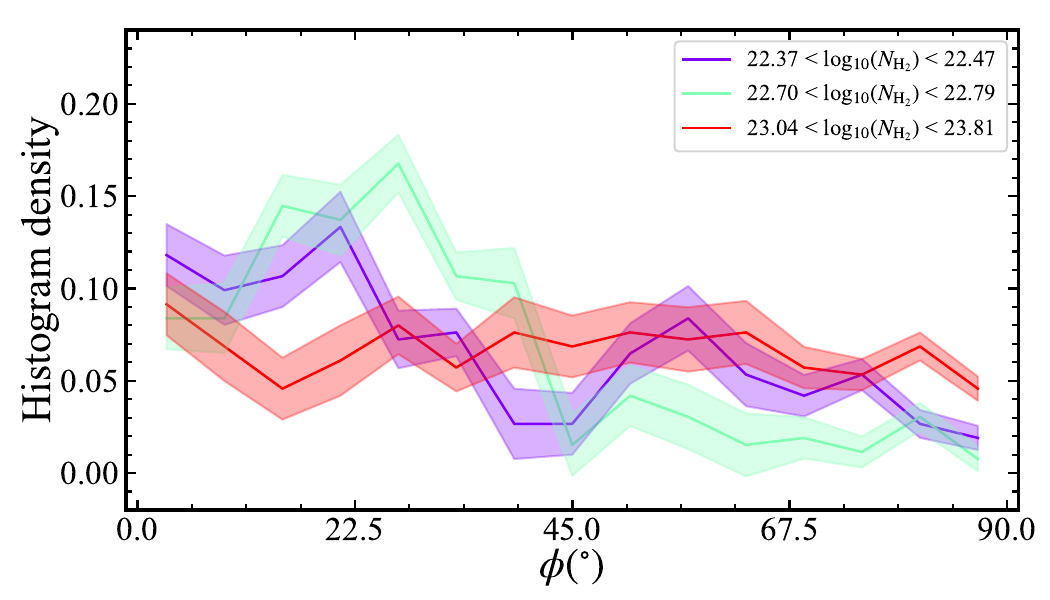}~
\includegraphics[width=0.32\linewidth]{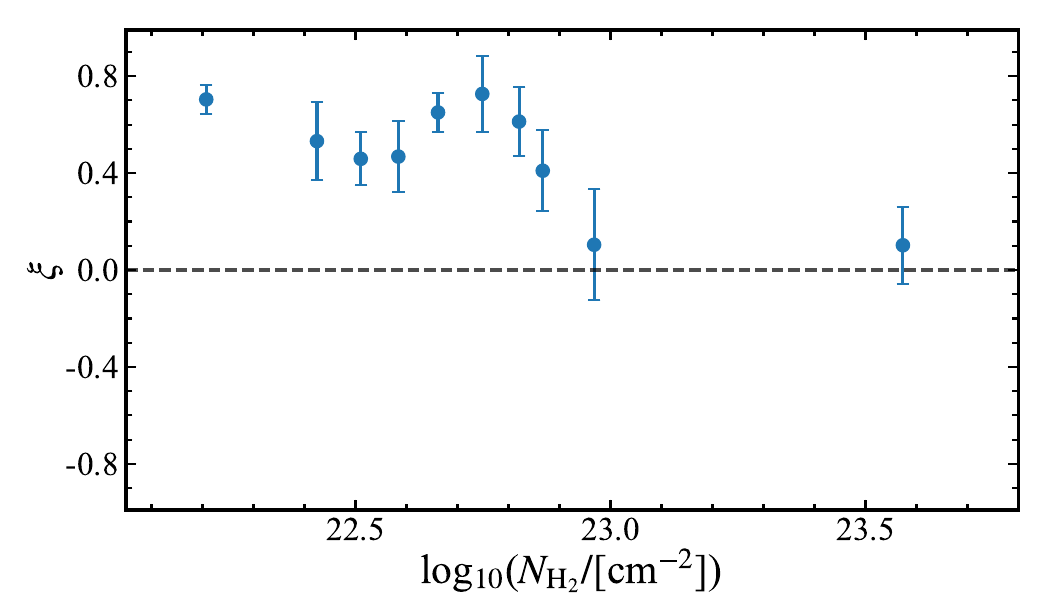}\\
\vspace{0.21cm}
\includegraphics[width=0.32\linewidth]{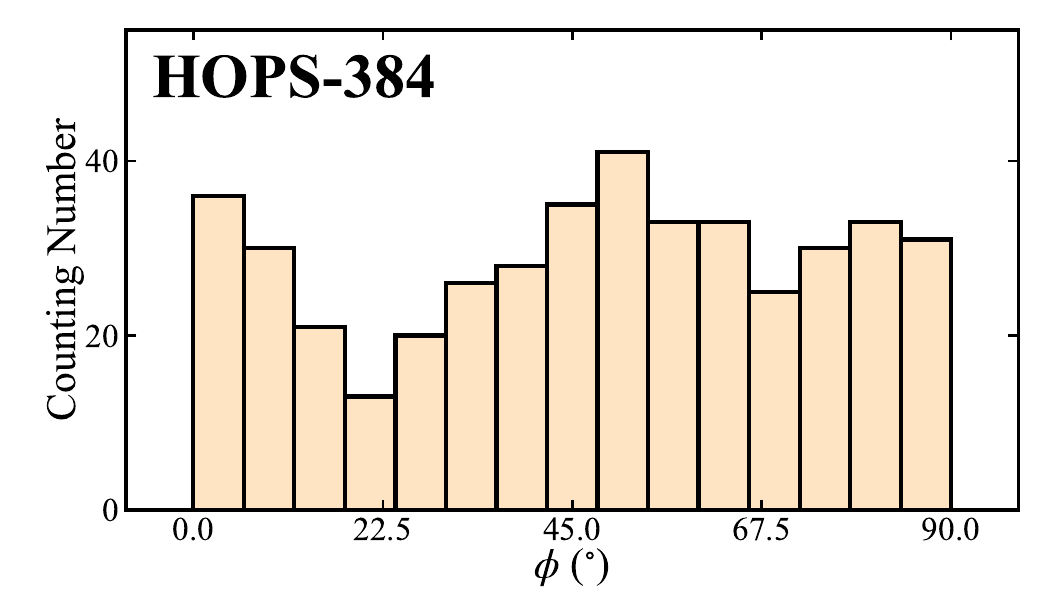}~
\includegraphics[width=0.32\linewidth]{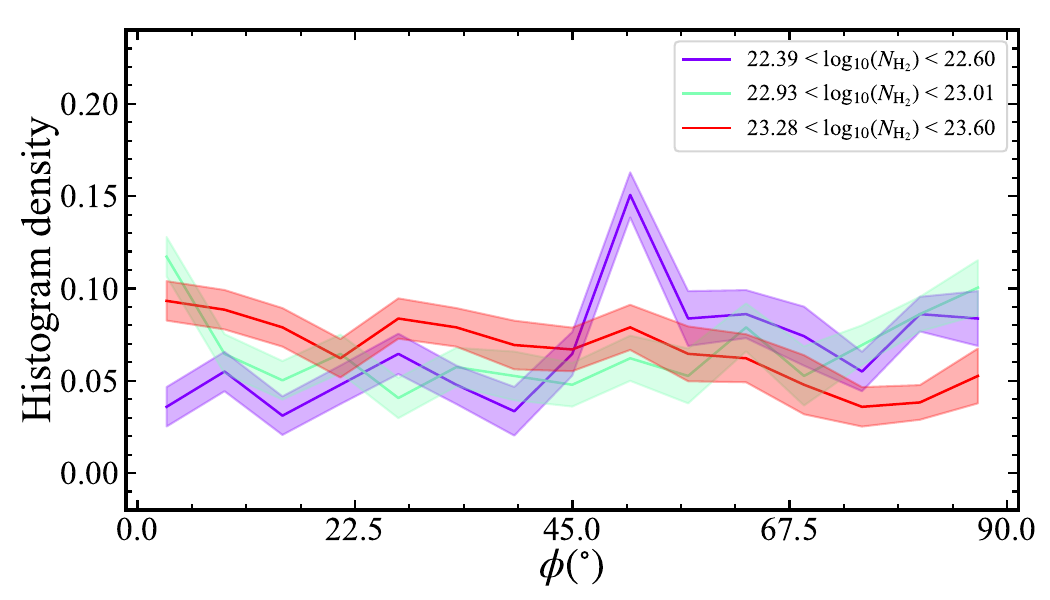}~
\includegraphics[width=0.32\linewidth]{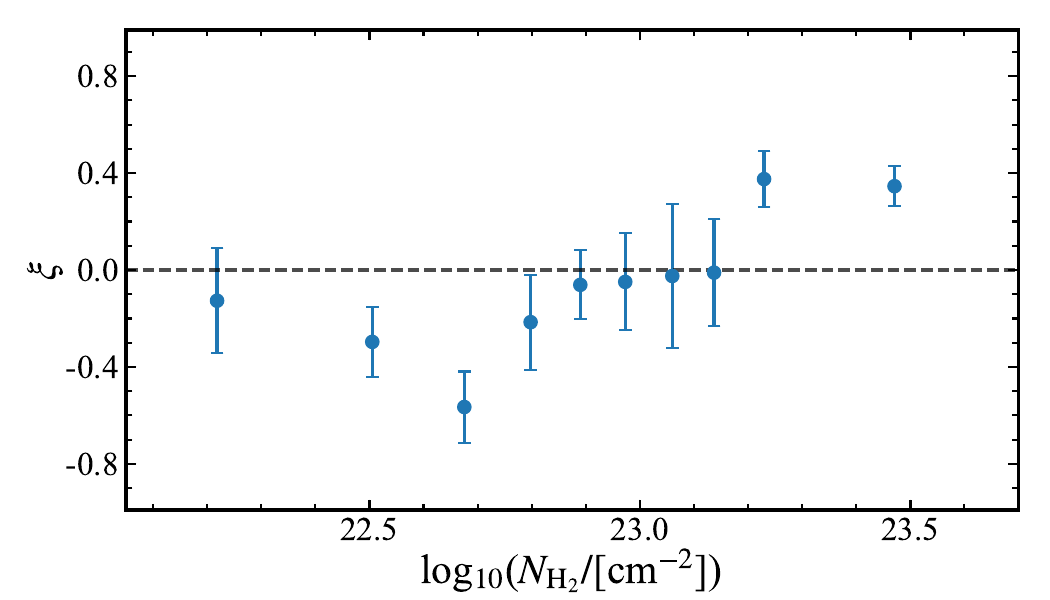}\\
\text{\textbf{{Figure 3.}} Histograms of relative orientation (HROs) between the $N_{\rm H_{2}}$ contours and the {\em B}-field orientation (Continued.)}
\end{figure*}

\begin{figure*}
\centering 
\includegraphics[width=0.32\linewidth]{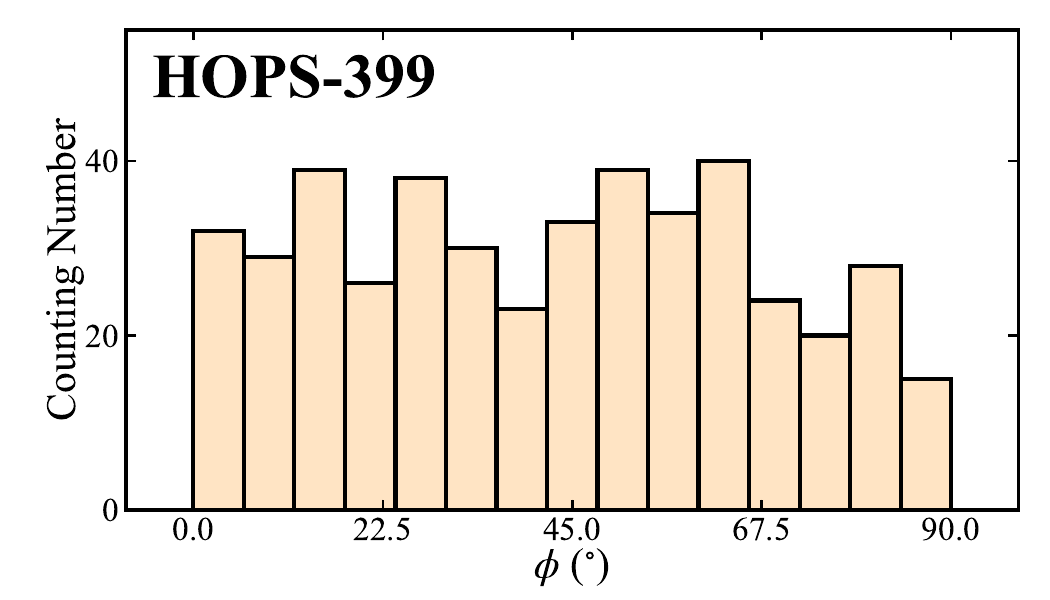}~
\includegraphics[width=0.32\linewidth]{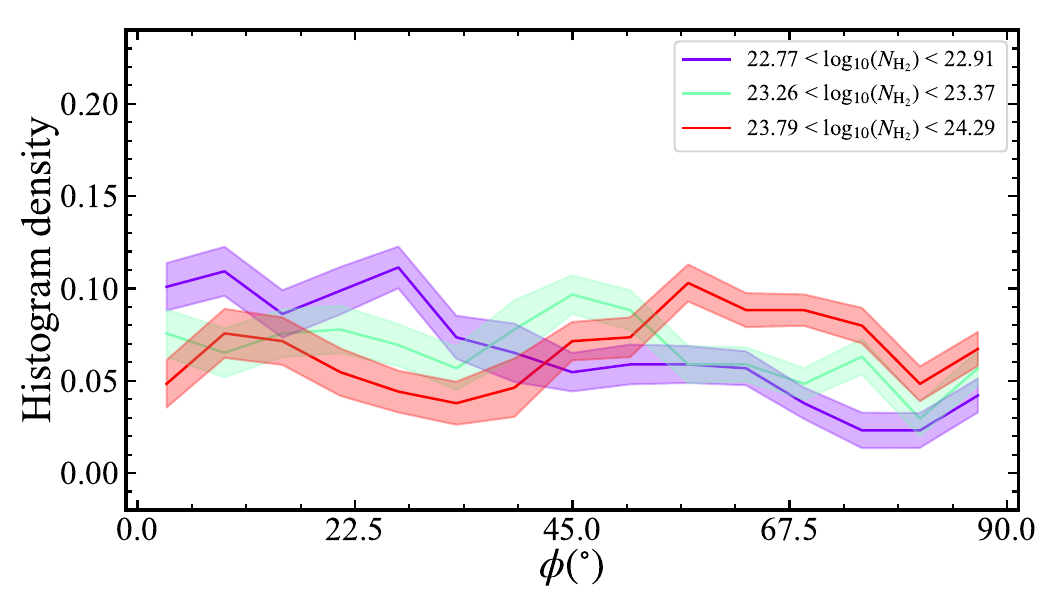}~
\includegraphics[width=0.32\linewidth]{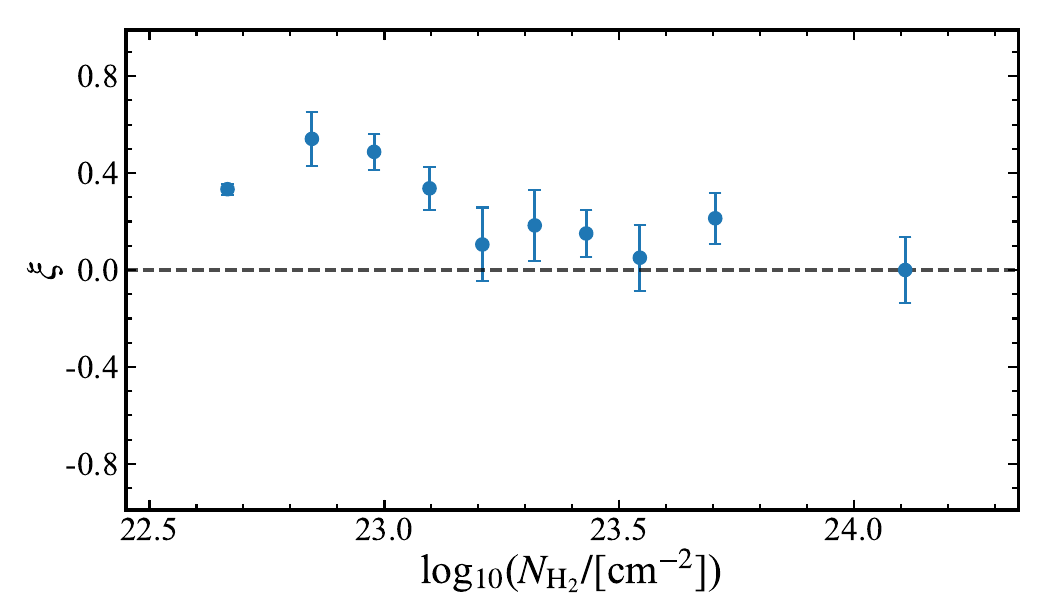}\\
\vspace{0.21cm}
\includegraphics[width=0.32\linewidth]{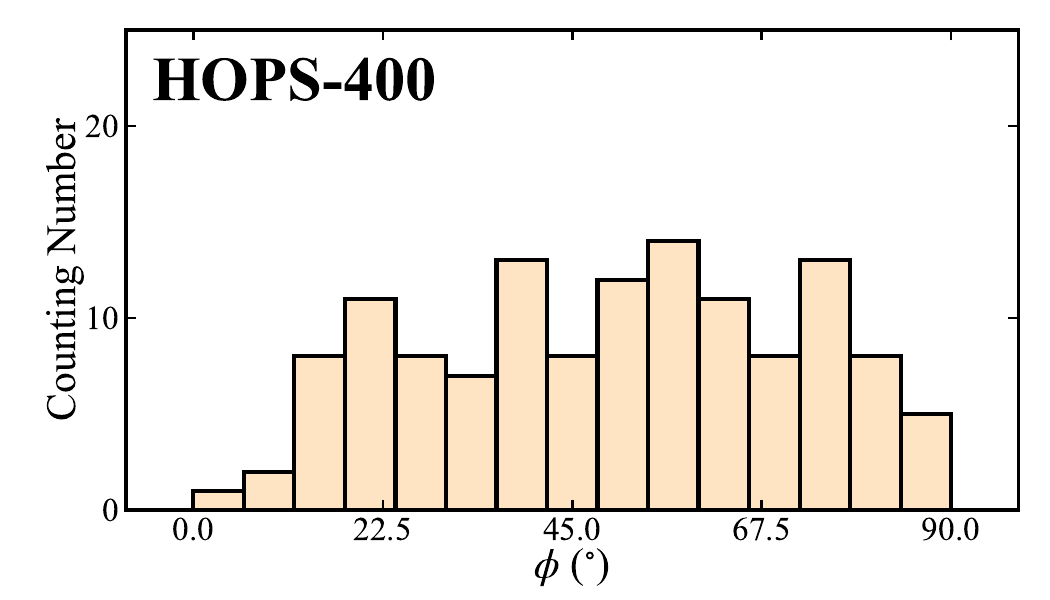}~
\includegraphics[width=0.32\linewidth]{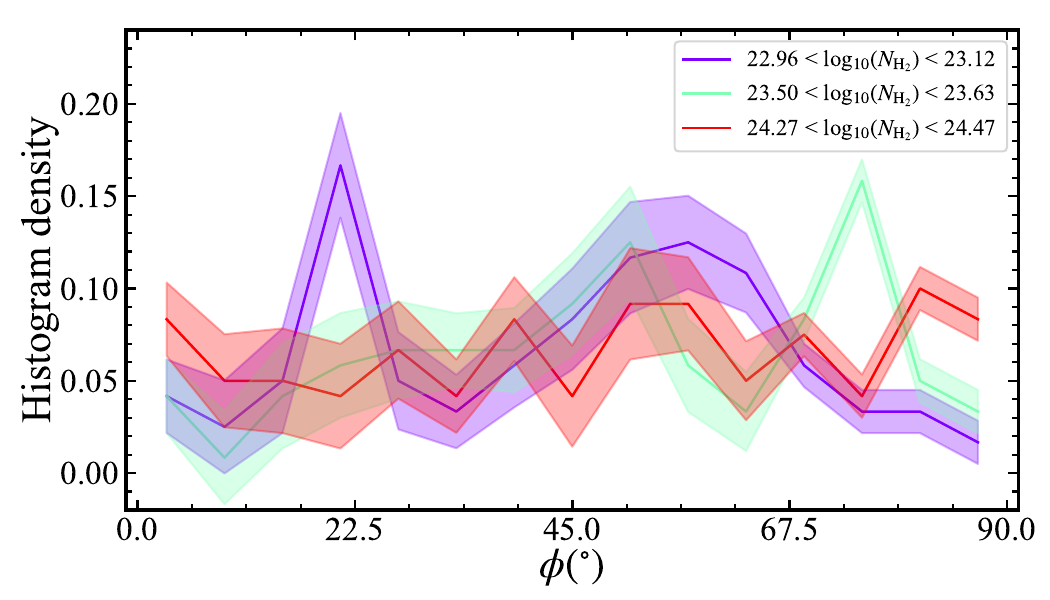}~
\includegraphics[width=0.32\linewidth]{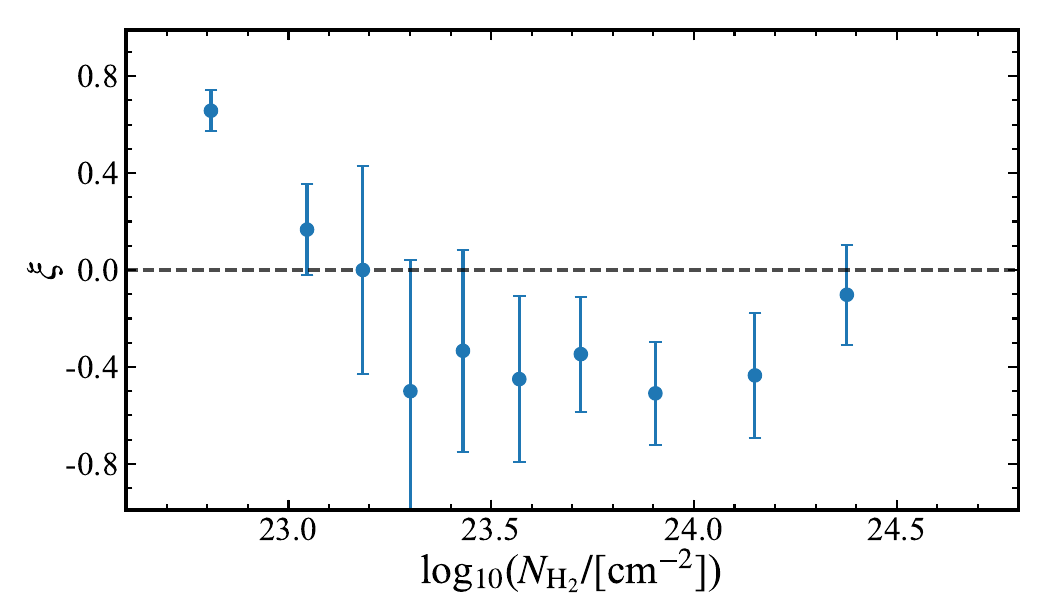}\\
\caption{Histograms of relative orientation (HROs) between the $N_{\rm H_{2}}$ contours and the {\em B}-field orientation. For each source, three panels are presented horizontally. 
{\it Left panel:} Global histograms of relative alignment $\phi$, with bin number of 15.
{\it Middle panel:} HROs for three distinct $N_{\rm H_{2}}$ ranges. The figures present the HROs for the lowest bin, an intermediate bin, and the highest $N_{\rm H_{2}}$ bin (purple, green, and red, respectively).
These bins have equal numbers of selected pixels within the indicated $N_{\rm H_{2}}$ ranges.
{\it Right panel:} Relative orientation parameter $\xi$ calculated for the different $N_{\rm H_{2}}$ bins.
The values $\xi>0$ and $\xi<0$ correspond to the {\em B}-field being oriented mostly parallel or perpendicular to the $N_{\rm H_{2}}$ contours, respectively. 
The dashed line is $\xi=0$, which corresponds to the case where there is no preferred relative orientation.}   
\label{fig:hro1} 
\end{figure*}

In the following analysis, we utilized the original pixel size ($\sim0.1''$), with the typical synthesized beam FWHM of $\sim0.8''$ in the observations, ensuring adequate sampling of the derivatives in each case \citep{Soler2017}.
We applied the HRO method to systematically analyze the relationship between the {\em B}-field orientation and the density structure. 
The relative orientation angle ($\phi$) between the tangent to the local density contours and the {\em B}-field is defined as
\begin{equation}\label{eq3}
\phi=\arctan\biggl(\frac{\vert\boldsymbol{\nabla N_{\rm H_{2}}\times\hat{E}_{p}}\vert}{\nabla N_{\rm H_{2}}\cdot\boldsymbol{\hat{E}_{p}}}\biggr),
\end{equation}
where $\boldsymbol{\hat{E}_{p}}$ is the unit polarization pseudo-vector (from which the {\em B}-field orientation is inferred by $90^{\circ}$ rotation) defined by the polarization angle $\theta_{p}$, and $\boldsymbol{\nabla N_{\rm H_{2}}}$ is the column density gradient.
The relative orientation $\phi$ is restricted to the range $0^{\circ}$--$90^{\circ}$, with angles outside this range transformed into equivalent values. We emphasize that this analysis is limited to the projected 2D geometry on the plane of the sky (POS).  The intrinsic 3D distribution of the gas and the inclination of the {\em B}-field are not accessible.

To estimate the column density gradient, we followed the method of \cite{sokolov2019structure} and \cite{jiao2024structure}.
The column density distribution within each $3\times3$ pixel grid was approximated by a first-degree bivariate polynomial,
\begin{equation}\label{eq4}
f(\alpha,~\delta)=N_{0}+a\Delta\alpha+b\Delta\delta,
\end{equation}
where $N_{0}$ is the column density in the central pixel of the grid, and $\Delta\alpha$ and $\Delta\delta$ are the pixel offsets along the right ascension and declination, respectively.
Only pixels with signal-to-noise ratio greater than 5$\sigma$ in dust emission were included, and we required at least eight valid neighboring pixels within the $3\times3$ grid for reliable gradient estimates. 
The gradient coefficients $(a,b)$ were determined by a least-squares fit,
\begin{equation}\label{eq5}
p(r) = \mathop{\rm argmin}_{a,b} 
\sum_{\Delta\alpha,\Delta\delta \leq 1} \biggl(N_{\rm H_{2}}(r^{\prime}) - f(r^{\prime},a,b) \biggr)^2,
\end{equation}
where $r^{\prime}=(\alpha+\Delta\alpha,~\delta+\Delta\delta)$ denotes the offset position, and the operator \texttt{argmin} returns the parameter values of $a$ and $b$ that minimize the given function.
The gradient orientation $\nabla N_{\rm H_{2}}$ and its uncertainty $\sigma$ are then given by
\begin{equation}\label{eq6}
\nabla N_{\rm H_{2}} = \arctan\frac{b}{a},
\end{equation}
\begin{equation}\label{eq7}
\sigma = \frac{1}{a^{2}+b^{2}}\sqrt{a^{2}\sigma_{b}^{2}+b^{2}\sigma_{a}^{2}},
\end{equation}
where $\sigma_{a}$ and $\sigma_{b}$ are the uncertainties in the fitted coefficients \citep{planck2016XXXV}.

Figure~\ref{fig:angle_difference} shows the spatial distribution of the relative alignment angle $\phi$ between the density contours and the {\em B}-field orientation, while the corresponding histograms for eight protostars are presented in the left panels of Figure~\ref{fig:hro1}.
In HOPS-182, HOPS-361, and HOPS-370, large portions of the envelopes are dominated by small $\phi$ values, with histograms skewed toward $\phi\lesssim30^{\circ}$, indicating widespread alignment between the density structures and the {\em B}-fields.
By contrast, HOPS-87 and HOPS-400 display extended regions with large $\phi$ values and histograms skewed toward $\phi\sim50^{\circ}$--$80^{\circ}$, suggesting a predominantly perpendicular tendency.
The remaining sources (HOPS-359, HOPS-384 and HOPS-399) show more mixed distributions without a clear preference. 

To quantify the dependence of these relative orientations on column density, following \cite{Soler2013} and \cite{Soler2017}, we construct the HROs (see the right panels of Figure~\ref{fig:hro1} for each source) by dividing the valid pixels of each source into ten column density intervals, each containing an equal number of pixels to ensure comparable statistics.
For each interval, we constructed histograms of the relative orientation angle $\phi$ between the polarization orientation ($\theta_{p}$) and the density gradient orientation ($\nabla N_{\rm H}$), i.e., between the inferred {\em B}-field orientation and the local density contour orientation.
For each source, the middle panels of Figure~\ref{fig:hro1} show the representative HROs for the lowest (blue), intermediate (green) and highest (red) column density intervals.

\subsection{Histogram shape parameter \texorpdfstring{$\xi$}{xi}} \label{subsec:3.3}

The histogram shape parameter $\xi$ quantitatively characterizes the shape of each HRO and is defined as
\begin{equation}\label{eq10}
\xi = \frac{A_{0} - A_{90}}{A_{0} + A_{90}},
\end{equation}
where $A_{0}$ is the area under the histogram for $0^{\circ} < \phi < 22.5^{\circ}$, and $A_{90}$ is the area under the histogram for $67.5^{\circ} < \phi < 90^{\circ}$. 
The parameter $\xi$ is nearly independent of the number of bins used to construct the histogram if the bin widths are smaller than the integration range. In our analysis, we calculated $\xi$ across 10 column density intervals. For each interval, the histograms were constructed with 15 bins, which corresponds to the bin width of $6.0^{\circ}$, significantly smaller than the integration range of $22.5^{\circ}$.

The variance of $\xi$ is given by
\begin{equation}\label{eq11}
\sigma_{\xi}^{2}=\frac{4(A_{90}^{2}\sigma_{A_{0}}^{2}+A_{0}^{2}\sigma_{A_{90}}^{2})}{(A_{0}+A_{90})^{4}},
\end{equation}
where $\sigma_{A_{0}}^{2}$ and $\sigma_{A_{90}}^{2}$ are the variances of the counts in the corresponding angle ranges, which represent the statistical ``jitter" of the histograms.
If the jitter is large such that $\sigma_{\xi}^{2}\gtrsim |\xi|$, the relative orientation becomes indeterminate. 
While the jitter depends on the chosen number of bins, the value of $\xi$ itself does not.
By construction, $\xi > 0$ indicates that the {\em B}-field tends to align parallel to the density structure contours, while $\xi < 0$ indicates that the {\em B}-field is preferentially perpendicular to the contours.
The right panels of Figure~\ref{fig:hro1} show $\xi$ as a function of $\log_{10}(N_{\rm H_{2}})$ for each column density interval, in each source.

\subsection{Structure Function}\label{subsec:3.4}

\begin{figure*} 
\centering 
\textbf{~~~~~~~~~~~~~HOPS-87~~~~~~~~~~~~~~~~~~~~~~~~~~~~~~~~~~~~~~~~~~~~~~~~~~~~~~~~~~~~~~~~~~~~~~~~~~~~~~~~~~~~~~~~~~~~~~~~~~~~~~~~HOPS-182}\\
\includegraphics[width=0.49\linewidth]{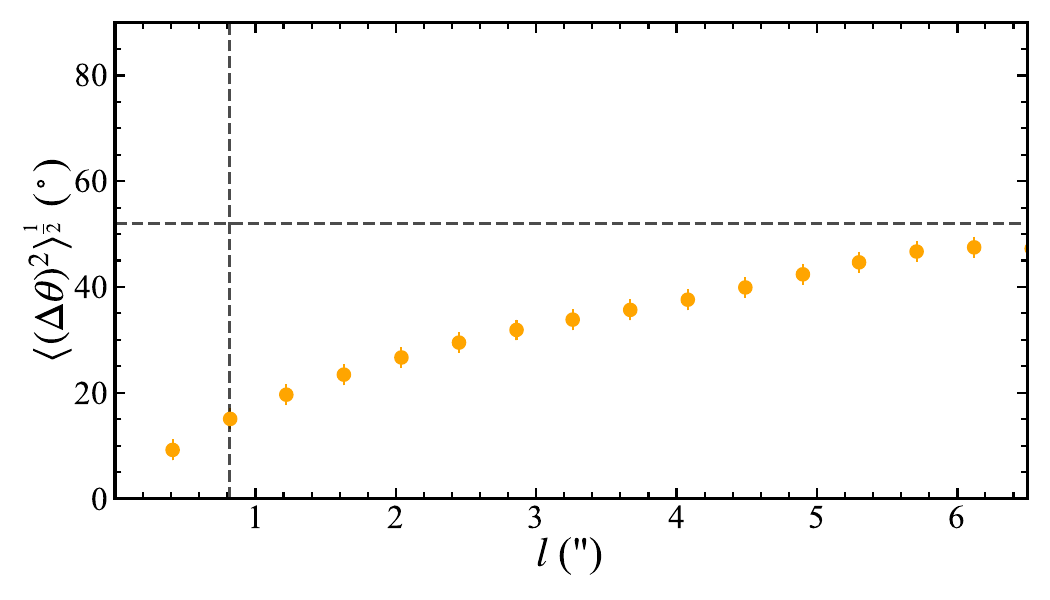}~~
\includegraphics[width=0.49\linewidth]{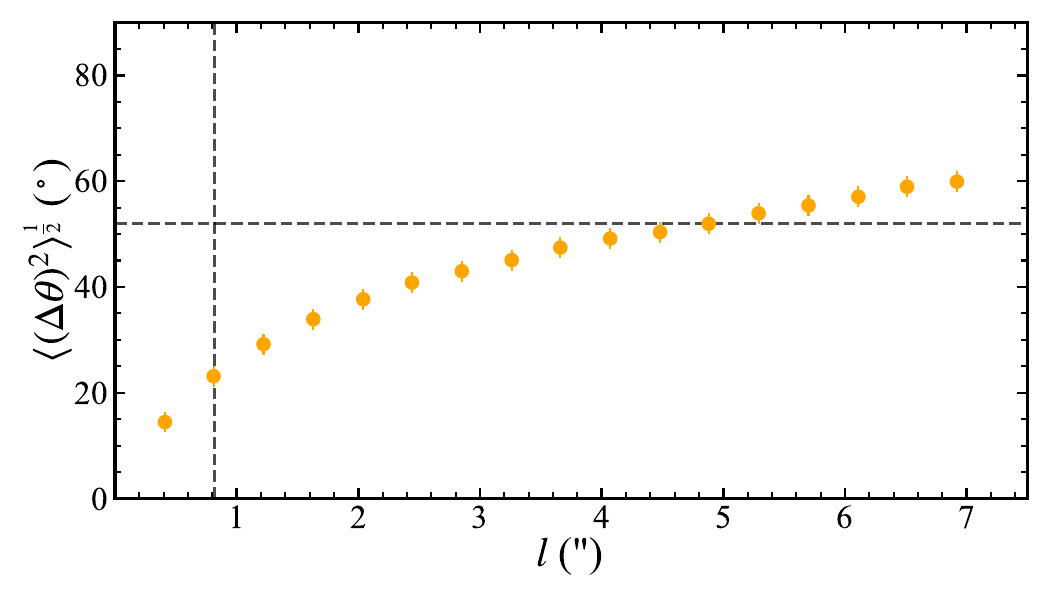}\\
\textbf{~~~~~~~~~~~~~HOPS-359~~~~~~~~~~~~~~~~~~~~~~~~~~~~~~~~~~~~~~~~~~~~~~~~~~~~~~~~~~~~~~~~~~~~~~~~~~~~~~~~~~~~~~~~~~~~~~~~~~~~~~HOPS-361}
\includegraphics[width=0.49\linewidth]{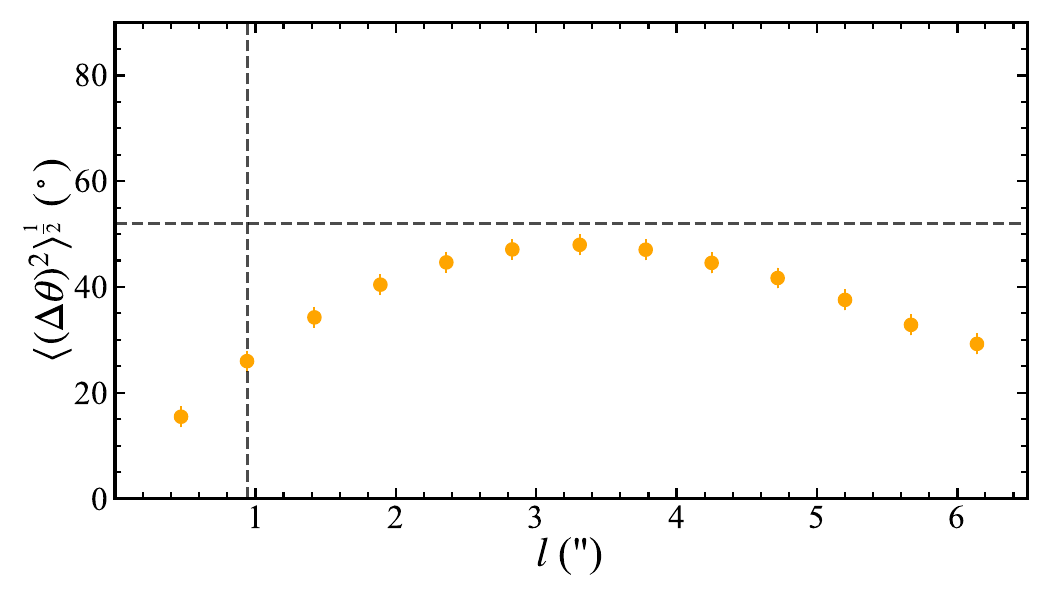}~~
\includegraphics[width=0.49\linewidth]{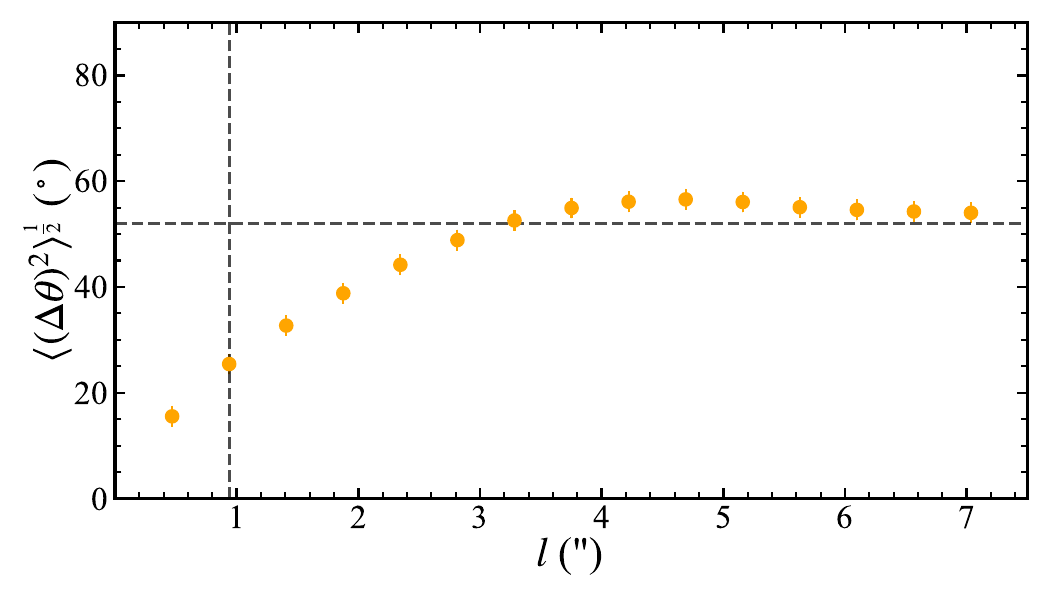}\\
\textbf{~~~~~~~~~~~~~HOPS-370~~~~~~~~~~~~~~~~~~~~~~~~~~~~~~~~~~~~~~~~~~~~~~~~~~~~~~~~~~~~~~~~~~~~~~~~~~~~~~~~~~~~~~~~~~~~~~~~~~~~~~HOPS-384}\\
\includegraphics[width=0.49\linewidth]{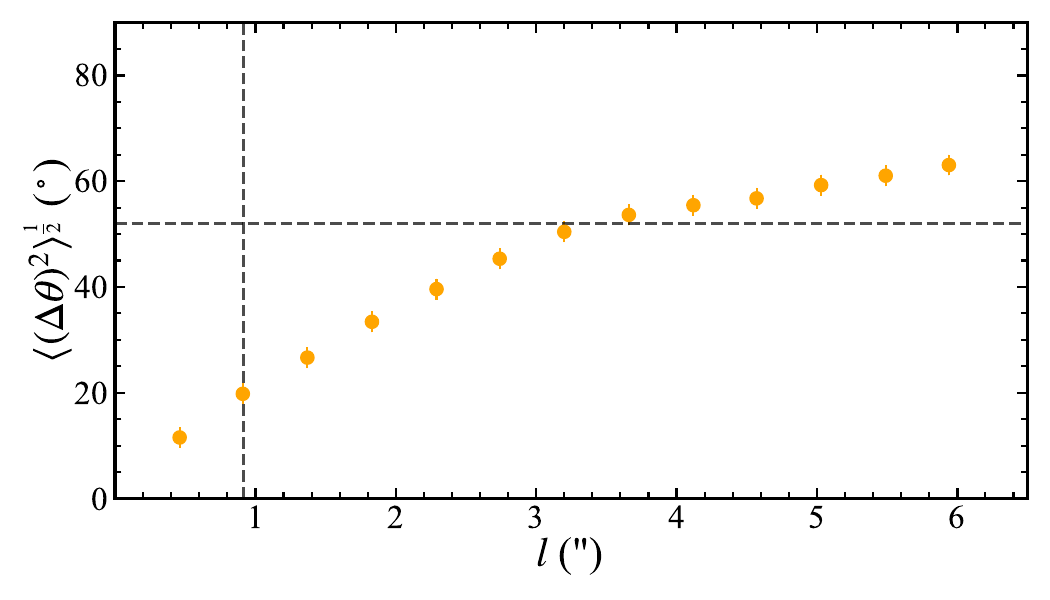}~~ 
\includegraphics[width=0.49\linewidth]{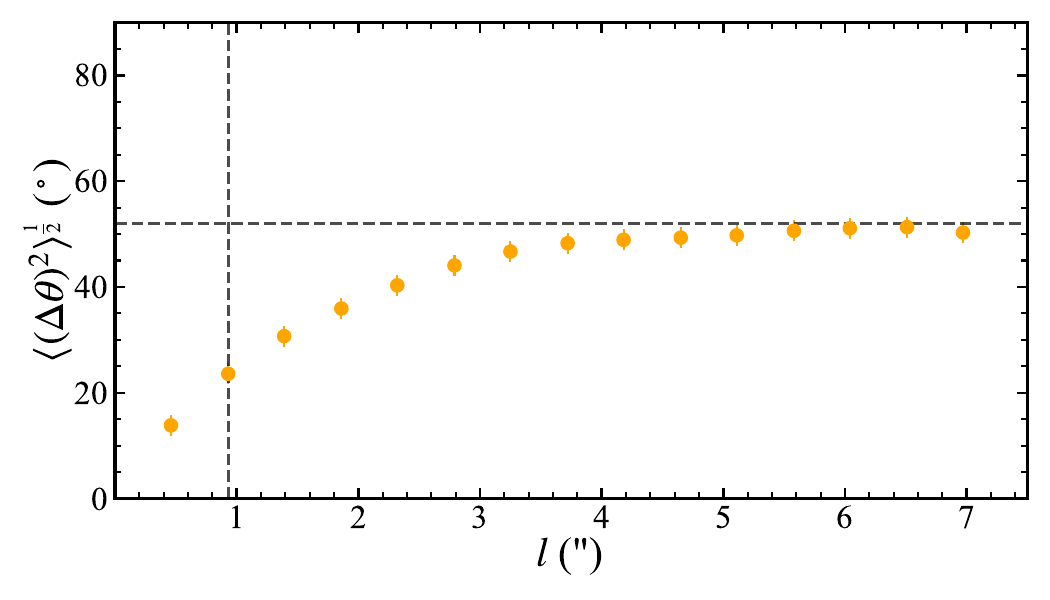}\\
\textbf{~~~~~~~~~~~~~HOPS-399~~~~~~~~~~~~~~~~~~~~~~~~~~~~~~~~~~~~~~~~~~~~~~~~~~~~~~~~~~~~~~~~~~~~~~~~~~~~~~~~~~~~~~~~~~~~~~~~~~~~~~HOPS-400}\\
\includegraphics[width=0.49\linewidth]{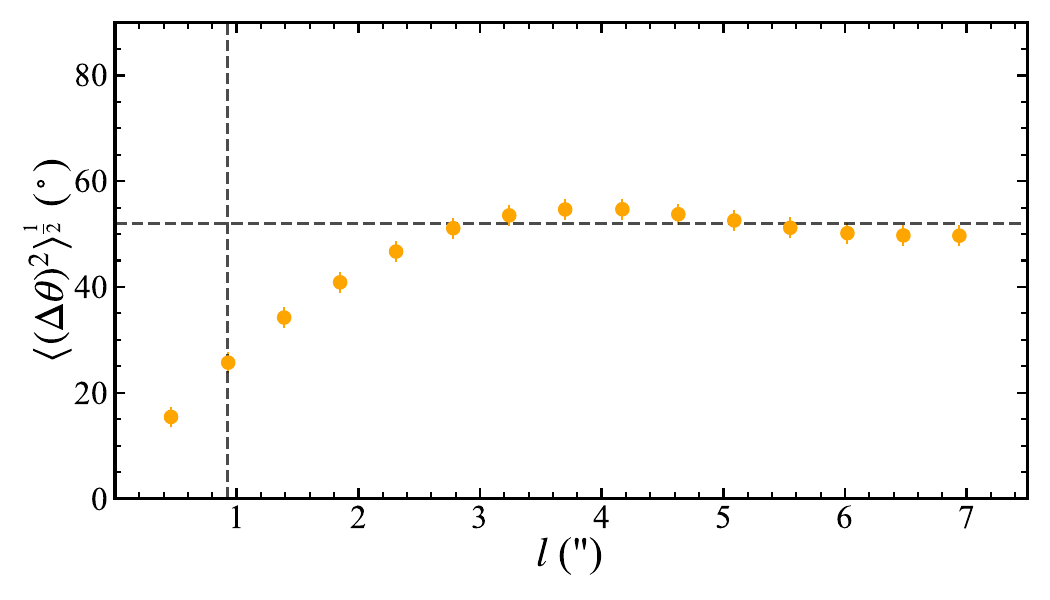}~~ 
\includegraphics[width=0.49\linewidth]{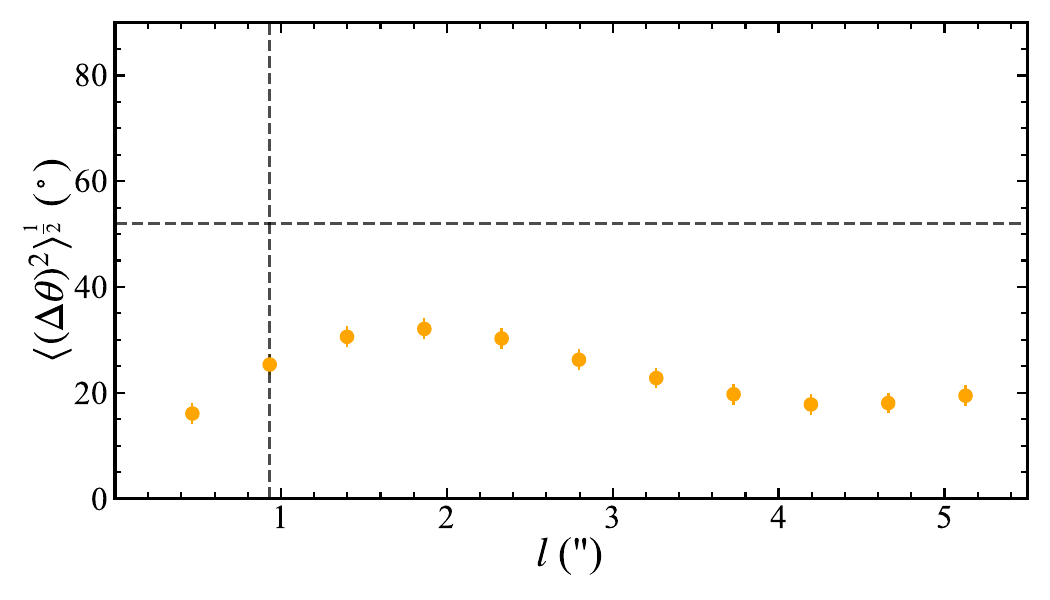}\\
\caption{Structure function of the {\em B}-field angles.
The horizontal dashed line at $52^\circ$ indicates the expected value for a purely random distribution, and the vertical dashed line indicates the beam size.}
\label{fig:sf}      
\end{figure*}

To characterize the {\em B}-field properties, we utilize the angular dispersion structure function (hereafter SF). 
This analysis is rooted in the Davis-Chandrasekhar-Fermi (DCF) technique \citep{Davis1951, Chandrasekhar+Fermi1953}, which links the dispersion of polarization angles to the ratio of turbulent-to-ordered magnetic energy. 
However, unlike the classical DCF method which typically relies on a single global dispersion estimate, we adopt the generalized formulation developed by \cite{Houde2009II}.
This approach analyzes the difference in polarization angles as a function of spatial separation. 
A key advantage of this formulation is its ability to account for signal averaging along the line of sight. 
Since the observed polarization is an integration of emission from multiple turbulent cells along the column and within the beam, the measured dispersion is generally lower than the intrinsic 3D dispersion. 
This framework accounts for this effect by modeling the number of turbulent cells involved in the integration, thereby providing a more robust separation of the large-scale ordered {\em B}-field from the turbulent component \citep{Houde2009II, Houde2016IV}.
The small structure function amplitude indicates a well-ordered, coherent {\em B}-field, while a larger amplitude suggests stronger dispersion.
The second-order structure function of the
polarization angles $\langle\Delta\theta^{2}(l)\rangle^{1/2}$ is defined as the average of the squared difference between the polarization angles measured for all pairs of points separated by a distance $l$,
\begin{equation}
\langle\Delta\theta^{2}(l)\rangle^{1/2} = \langle\vert\theta(\boldsymbol r + \boldsymbol l)-\theta(\boldsymbol r)\vert^{2}\rangle^{1/2}~.
\end{equation}

SF-based implementations has been widely used as a powerful statistical tool to infer the relationship between the large-scale ordered and turbulent components of {\em B}-field in molecular clouds \citep[e.g.,][]{Houde2009II,Houde2016IV,Franco2010}. In this work, we restrict our analysis to polarization angle measurements with signal-to-noise ratios greater than 3$\sigma$ in polarized intensity $P$. 
The SF is computed using spatial separations sampled at half the beam size to ensure Nyquist sampling. The resulting structure function is shown in Figure~\ref{fig:sf}.

\section{Discussion} \label{sec:4}

The variation in HRO shapes and the $\xi$ parameter presented in Section \ref{sec:3} indicate complex interactions between {\em B}-fields and the envelope gas. In this Section, we discuss the physical mechanisms of relative alignment between the column density structure and {\em B}-fields, including the dependence on the column density and magnetization levels on the envelope scales, and we finally assess the projection effects and other factors.

\subsection{HRO analysis: Does column density affect the relative alignment?}
\label{subsec:4.1}

The middle panels of Figure~\ref{fig:hro1} show the HROs in different column-density bins for the sample.
In HOPS-182, HOPS-361, and HOPS-399, the HROs for the lowest and intermediate $N_{\rm H}$ ranges peak near $\phi\sim0^{\circ}$, with significantly more counts in $0^{\circ}<\phi<22.5^{\circ}$ than in $67.5^{\circ}<\phi<90^{\circ}$.
At the highest $N_{\rm H}$, the distributions shift toward larger angles ($\phi\sim50^{\circ}$--$80^{\circ}$).
HOPS-370 shows a similar trend, exhibiting alignment at low and intermediate densities but flatter HROs at high densities.
In contrast, HOPS-87 shows HROs weighted toward large angles at the highest column densities, while the lower-density bins peak at intermediate $\phi$ ($40^{\circ}$--$65^{\circ}$).
For HOPS-384 and HOPS-400, the lowest-density HROs peak at relatively small angles but flatten with increasing $N_{\rm H}$, whereas HOPS-359 shows the opposite behavior, with large-angle peaks at intermediate densities and flatter HROs at low and high densities.


Previous studies suggest that the relative orientation between density structures and {\em B}-fields depends on column density, with alignments typically transitioning from parallel to perpendicular as the density increases.
For example, Planck observations and follow-up HRO analyses in nearby molecular clouds ($\sim 0.4$--$1.4$ pc) typically find $X_{\rm HRO}$ values of $\log(N_{\rm H_{2}}/{\rm cm^{-2}})\sim21.7$--$22.5$ \citep{Planck2016XIX, Soler2017}.
In Orion A on $\sim0.6$ pc scales, \citet{jiao2024structure} reported similar values of $21.0$--$21.6$, but noted that the transition column density increases when probed on smaller scales, reaching $\log(N_{\rm H_{2}}/{\rm cm^{-2}})\sim22.9$--$23.2$ at $\sim0.04$~pc.
On even smaller scales of $\sim 0.03$ pc, \cite{kwon2022hro} found that {\em B}-fields are parallel to less-dense filaments but perpendicular to dense, star-forming ones; however, at the highest densities the orientation returns to parallel, likely due to field dragging by infall, implying no single universal transition density.
In our sample, the values of $\xi$, as shown in the column (4) and (5) of Table~\ref{tab:source} and the right panels of Figure~\ref{fig:hro1}, for HOPS-182, HOPS-361, HOPS-370, and HOPS-399 are predominantly positive (i.e., parallel), whereas HOPS-87 and 400 are predominantly negative (i.e., perpendicular). For the rest sources, $\xi$ for HOPS-359 fluctuates around zero and shows no clear trend across the explored column density range, while HOPS-384 presents a transition from perpendicular to parallel alignment. Consequently, despite these individual protostars exhibiting distinct behaviors, in general, we find no clear evidence for a parallel-to-perpendicular transition with increasing column density at spatial scales probed in our data.

\subsection{Does the magnetization level affect the relative alignment?}
\label{subsec:4.2}

Previous BOPS results \citep{Huang2024} suggested that protostars with standard hourglass {\em B}-fields exhibit small velocity gradients due to strong magnetic braking, whereas large velocity gradients tend to be found towards sources with rotated hourglass, spiral, or complex morphologies where the {\em B}-field is less dominant relative to gravity and rotation.
Recent numerical simulations \citep{nacho2024magnetic} further support this interpretation, showing that strongly magnetized envelopes 
are expected to develop standard hourglass morphologies, while weakly magnetized cases 
instead produce rotated hourglass configurations.
Spiral field morphologies, on the other hand, have been suggested as an additional channel for angular momentum redistribution at the disk-envelope interface \citep{Wang2022}.
From the SF calculated for these sources (see Figure~\ref{fig:sf} and column (6) of Table~\ref{tab:source}), we find the following:
HOPS-87 exhibits a small SF at the smallest separations, followed by a monotonic increase with spatial scale. 
In contrast, the SFs of HOPS-182, HOPS-359, and HOPS-400 are small at small separations and increase with increasing lag as local {\em B}-field perturbations decorrelate, but decrease again at the largest separations. 
This downturn is likely caused by the dominance of a large-scale ordered {\em B}-field.
The remaining sources show SFs that increase with spatial scale and approach a plateau at large separations.
Notably, HOPS-87 and HOPS-400 exhibit relatively small overall SF amplitudes ($<52^{\circ}$), with values remaining below those expected for a random field at all spatial scales.
The SF in HOPS-359 likewise stays below $52^{\circ}$, but reaches comparatively high values at the small separations ($\sim2.5^{\prime\prime}-4.5^{\prime\prime}$).
The other sources, however, display comparatively larger SF amplitudes over most spatial scales, indicating larger {\em B}-field dispersion. 
It is consistent with the velocity gradient analysis by \cite{Huang2024}, which suggests that the dispersion arises from dynamical processes such as gravitational collapse and angular momentum (e.g., rotational motions) that significantly twist the field lines at envelope scales.
We thus expect that HOPS-87 and HOPS-400 are more magnetized than the remaining sources.


Visual inspection of the LIC textures of the {\em B}-field patterns overlaid on the column density maps (right panels of Figure~\ref{fig:obs}) reveals that {\em B}-fields tend to align parallel to the elongated density structures in HOPS-182, HOPS-361, and HOPS-370.
This impression is consistent with the histograms in the left panels of Figure~\ref{fig:hro1}, which show that large portions of their envelopes are dominated by small relative angles.
By contrast, the more magnetized sources HOPS-87 and HOPS-400 display relative larger fraction of high angle values (i.e., perpendicular alignment).
From the results of the HRO analysis (right panels of Figure~\ref{fig:hro1}), most $\xi$ values in HOPS-87 and HOPS-400 are almost negative, suggesting a predominantly perpendicular configuration that becomes more pronounced across column density.
The less magnetized sources, on the other hand, exhibit predominantly positive $\xi$ values (HOPS-182, HOPS-361, HOPS-370, and HOPS-399) or show mixed positive and negative $\xi$ values (HOPS-359 and HOPS-384), suggesting a parallel alignment or a transitional behavior without a dominant alignment trend. 

Theoretical and observational studies show that the relative alignment between {\em B}-fields and column density structures correlates with the level of magnetization \citep[e.g.,][]{hull2017, kwon2022hro}.
In strongly magnetized, sub-Alfv{\'e}nic environments, {\em B}-fields tend to be well ordered and oriented parallel to the gas flow, as gravity and gas flows are guided along the field lines. 
In weakly magnetized or super-Alfv{\'e}nic regimes where gravity dominates, on the other hand, the field orientation becomes parallel or randomly aligned with the density structures.
Our results appear to show a consistency with previous studies that the magnetization level correlates with the relative alignment between density structures and {\em B}-fields on envelope scales, suggesting that magnetization plays a key role in determining the relative alignment between {\em B}-fields and density structures on envelope scales.

\subsection{Possible factors influencing the results}
\label{subsec:4.3}

\begin{figure} 
\centering 
\includegraphics[width=0.78\linewidth]{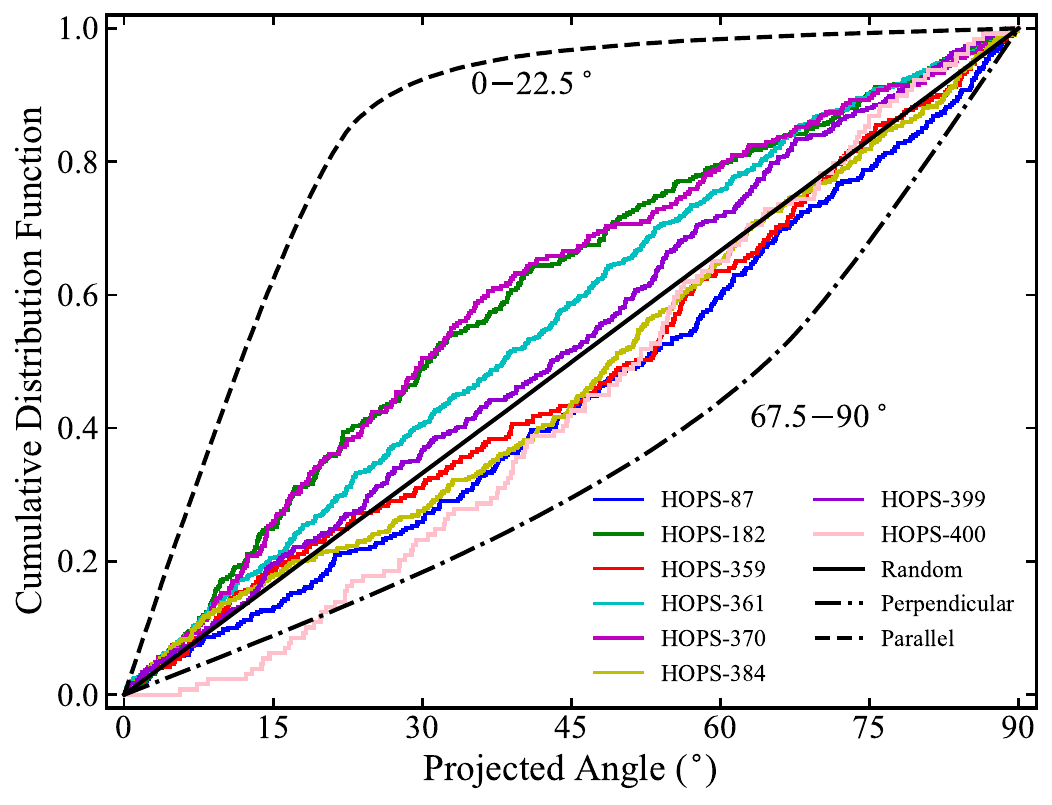}
\caption{Cumulative distribution functions of the projected relative angles.
The colored curves are the observed angle difference between the {\em B}-field orientation and the density contours for the eight sources.
The black solid line, dashed line, and dotted–dashed line indicate Monte Carlo simulations of the expected projected angle for two vectors that are 3D random (difference angle of two vectors is between $0.0^{\circ}$ and $90.0^{\circ}$), parallel ($0.0^{\circ}-22.5^{\circ}$), and perpendicular ($67.5^{\circ}-90.0^{\circ}$), respectively.}
\label{fig:cdf}      
\end{figure}

It is important to note that both the density distribution and the {\em B}-field morphology are inherently three-dimensional, whereas our observations capture only their two-dimensional projection onto the plane of the sky.
To assess the impact of projection effects, we constructed cumulative distribution functions (CDFs) of the observed angle differences $\phi$ and compared them with simulated distributions of 2D intersecting angles projected from a uniform distribution in 3D space.
The simulations considered three scenarios: (i) a random distribution, where 3D angles are uniformly sampled between $0^{\circ}$ and $90^{\circ}$ and then projected; (ii) a parallel case, with 3D angles restricted to $0^{\circ}$--$22.5^{\circ}$; and (iii) a perpendicular case, with 3D angles restricted to $67.5^{\circ}$--$90^{\circ}$.
As shown in Figure~\ref{fig:cdf}, the observed distributions for HOPS-182, HOPS-361 and HOPS-370 lie between the parallel and random models, indicating a tendency toward alignment but with appreciable scatter.
By contrast, the distribution for HOPS-87 and HOPS-400 lie between the perpendicular and random models, potentially suggesting a weak preference for perpendicularity.
The remaining sources (HOPS-359, HOPS-384, and HOPS-399) closely follow the random distribution.
Therefore, projection effects can broaden the observed distributions and introduce scatter in the measured $\xi$ values, but do not erase systematic differences among sources.

Outflows and related small-scale structures can also influence the measured relative orientations.
Bipolar outflows carve cavities into the envelope, sweeping up material and locally distorting both the density distribution and the {\em B}-field geometry.
In such cases, polarization detected along the cavity walls may not trace the collapse-driven field, but rather dust grains aligned by radiative anisotropies or shocks, potentially biasing the inferred alignment \citep[e.g.,][]{le2019b}.
Streamers, as likely seen in HOPS-182, HOPS-361 and HOPS-370 \citep{Huang2024}, which channel material from large scales onto the protostellar envelope or disk, can likewise drag field lines and impose coherent density features that mimic or obscure the collapse-driven orientation \citep[e.g.,][]{pineda2020streamer}.
At smaller radii, polarized emission may also arise from self-scattering rather than magnetically aligned grains, producing polarization patterns that are not related to the {\em B}-field orientation \citep[e.g.,][]{Kataoka2015, Yang2016, Stephens2017, Stephens2023Nature}.
However, this effect is likely negligible in our analysis, since we restricted the study to envelope scales, much larger than the disk scales where scattering dominates, and explicitly excluded those BOPS sources dominated by self-scattering \citep[see][]{Huang2024}.

In addition, uncertainties in the derivation of column densities, arising from assumptions about dust temperature, opacity, or gas-to-dust ratio, can shift the absolute values of $N_{\rm H}$ and thereby affect the determination of transition column densities and the inferred density-field alignment.
We also note that our results are based on only eight sources, which limits the statistical robustness of the observed trends. 
Despite these limitations, this study constitutes the first sample of objects studied at these scales ($\sim400$ au).
Expanding this analysis to larger, well-characterized samples will be essential to confirm the role of density and magnetization in controlling the relative orientation between density structures and {\em B}-fields.

\section{Conclusions} \label{sec:5}

In this work, we investigated the relative alignment between density structures and {\em B}-fields in BOPS protostellar envelopes on envelope scales, as this relationship provides an observational test of how {\em B}-fields regulate, or yield to, gravity during the earliest stages of star formation.
Column density maps were derived from 870~$\mu$m dust continuum emission, and the HRO method was applied to quantify the relative alignment between {\em B}-fields and density structures. Unlike molecular cloud scales, which $\xi$ present a parallel-to-perpendicular transition with $N_{\rm H}$ \citep[e.g.,][]{Planck2016XIX}, the $\xi$ in our sample shows overall little systematic evolution, implying that column density alone does not fully govern the density-field alignment for these sources.
On the other hand, weakly magnetized envelopes exhibit predominantly parallel or random alignment, while strongly magnetized ones appear to show mainly perpendicular configurations, suggesting that magnetization level may play a key role in determining the relative alignment between {\em B}-fields and density structures on envelope scales.

Several effects may influence the relative orientations measured on envelope scales, including projection effects, outflow cavities, streamers, and uncertainties in the column density derivation. 
Nevertheless, the observed differences among sources provide observational hints in highlighting the role of magnetization in shaping density–field alignment.
Overall, our results provide hints that the relative orientation between density structures and {\em B}-fields not only depends on its density, but may also depends on the level of magnetization. 
Expanding this analysis to larger samples will be essential to establish the statistical significance of these trends and to further constrain the role of {\em B}-fields in protostellar collapse.

\nolinenumbers
\begin{acknowledgments}
B.H., J.M.G. and A.S.-M. acknowledge support by the grant PID2020-117710GB-I00 and PID2023-146675NB-I00 (MCI-AEI-FEDER, UE).
Z.J.L. acknowledges support by NSFC grant No.12563005, 12494571 and the National Key Research and Development (R$\&$D) Program of China No.2024YFA1611704. 
A.S.-M. acknowledges support from the RyC2021-032892-I grant funded by MCIN/AEI/10.13039/501100011033 and by the European Union `Next GenerationEU’/PRTR.
This research is supported by the program Unidad de Excelencia María de Maeztu CEX2020-001058-M, by Guangxi Natural Science Foundation under Grant No.AD23026127, 2024GXNSFBA010436, the Guangxi Talent Program (``Highland of Innovation Talents”) and the Guangxi Qingmiao Talent Support Program, and also by the Korea Astronomy and Space Science Institute under the R\&D program (Project No. 2026-1-844-00) supervised by the Ministry of Science and ICT.
This paper makes use of the following ALMA data: ADS/JAO. ALMA\#2019.1.00086. ALMA is a partnership of ESO (representing its member states), NSF (USA) and NINS (Japan), together with NRC (Canada), MOST and ASIAA (Taiwan), and KASI (Republic of Korea), in cooperation with the Republic of Chile. The Joint ALMA Observatory is operated by ESO, AUI/NRAO, and NAOJ.
\end{acknowledgments}


\begin{thebibliography}{}
\expandafter\ifx\csname natexlab\endcsname\relax\def\natexlab#1{#1}\fi
\providecommand{\url}[1]{\href{#1}{#1}}
\providecommand{\dodoi}[1]{doi:~\href{http://doi.org/#1}{\nolinkurl{#1}}}
\providecommand{\doeprint}[1]{\href{http://ascl.net/#1}{\nolinkurl{http://ascl.net/#1}}}
\providecommand{\doarXiv}[1]{\href{https://arxiv.org/abs/#1}{\nolinkurl{https://arxiv.org/abs/#1}}}

\bibitem[{{A{\~n}ez-L{\'o}pez} {et~al.}(2024){A{\~n}ez-L{\'o}pez},
  {Lebreuilly}, {Maury}, \& {Hennebelle}}]{nacho2024magnetic}
{A{\~n}ez-L{\'o}pez}, N., {Lebreuilly}, U., {Maury}, A., \& {Hennebelle}, P.
  2024, \aap, 687, A63, \dodoi{10.1051/0004-6361/202245029}

\bibitem[{{Allen} {et~al.}(2003){Allen}, {Li}, \& {Shu}}]{Allen2003}
{Allen}, A., {Li}, Z.-Y., \& {Shu}, F.~H. 2003, \apj, 599, 363,
  \dodoi{10.1086/379243}

\bibitem[{Andersson {et~al.}(2015)Andersson, Lazarian, \&
  Vaillancourt}]{Andersson2015}
Andersson, B.-G., Lazarian, A., \& Vaillancourt, J.~E. 2015, Annual Review of
  Astronomy and Astrophysics, 53, 501.
\newblock \url{https://api.semanticscholar.org/CorpusID:123384442}

\bibitem[{{Barnes} {et~al.}(2025){Barnes}, {Ryder}, {Novak}, \&
  {Fissel}}]{Barnes2025}
{Barnes}, P.~J., {Ryder}, S.~D., {Novak}, G., \& {Fissel}, L.~M. 2025, \apj,
  985, 186, \dodoi{10.3847/1538-4357/adcedf}

\bibitem[{Cabral \& Leedom(1993)}]{cabral1993special}
Cabral, B., \& Leedom, L. 1993, Special Interest Group on GRAPHics and
  Interactive Techniques Proceedings, 263

\bibitem[{{Chandrasekhar} \& {Fermi}(1953)}]{Chandrasekhar+Fermi1953}
{Chandrasekhar}, S., \& {Fermi}, E. 1953, \apj, 118, 113,
  \dodoi{10.1086/145731}

\bibitem[{{Clark} {et~al.}(2014){Clark}, {Peek}, \& {Putman}}]{Clark2014}
{Clark}, S.~E., {Peek}, J.~E.~G., \& {Putman}, M.~E. 2014, \apj, 789, 82,
  \dodoi{10.1088/0004-637X/789/1/82}

\bibitem[{Dapp \& Basu(2010)}]{dapp2010}
Dapp, W.~B., \& Basu, S. 2010, Astronomy \& Astrophysics, 521, L56

\bibitem[{{Davis}(1951)}]{Davis1951}
{Davis}, L. 1951, Physical Review, 81, 890, \dodoi{10.1103/PhysRev.81.890.2}

\bibitem[{{Franco} {et~al.}(2010){Franco}, {Alves}, \& {Girart}}]{Franco2010}
{Franco}, G.~A.~P., {Alves}, F.~O., \& {Girart}, J.~M. 2010, \apj, 723, 146,
  \dodoi{10.1088/0004-637X/723/1/146}

\bibitem[{{Furlan} {et~al.}(2016){Furlan}, {Fischer}, {Ali}, {Stutz}, {Stanke},
  {Tobin}, {Megeath}, {Osorio}, {Hartmann}, {Calvet}, {Poteet}, {Booker},
  {Manoj}, {Watson}, \& {Allen}}]{furlan2016herschel}
{Furlan}, E., {Fischer}, W.~J., {Ali}, B., {et~al.} 2016, \apjs, 224, 5,
  \dodoi{10.3847/0067-0049/224/1/5}

\bibitem[{{Galli} \& {Shu}(1993{\natexlab{a}})}]{Galli1993a}
{Galli}, D., \& {Shu}, F.~H. 1993{\natexlab{a}}, \apj, 417, 220,
  \dodoi{10.1086/173305}

\bibitem[{{Galli} \& {Shu}(1993{\natexlab{b}})}]{Galli1993b}
---. 1993{\natexlab{b}}, \apj, 417, 243, \dodoi{10.1086/173306}

\bibitem[{{Girart} {et~al.}(2009){Girart}, {Beltr{\'a}n}, {Zhang}, {Rao}, \&
  {Estalella}}]{Girart2009}
{Girart}, J.~M., {Beltr{\'a}n}, M.~T., {Zhang}, Q., {Rao}, R., \& {Estalella},
  R. 2009, Science, 324, 1408, \dodoi{10.1126/science.1171807}

\bibitem[{{Girart} {et~al.}(1999){Girart}, {Crutcher}, \&
  {Rao}}]{girart1999detection}
{Girart}, J.~M., {Crutcher}, R.~M., \& {Rao}, R. 1999, \apjl, 525, L109,
  \dodoi{10.1086/312345}

\bibitem[{{Girart} {et~al.}(2013){Girart}, {Frau}, {Zhang}, {Koch}, {Qiu},
  {Tang}, {Lai}, \& {Ho}}]{girart2013dr}
{Girart}, J.~M., {Frau}, P., {Zhang}, Q., {et~al.} 2013, \apj, 772, 69,
  \dodoi{10.1088/0004-637X/772/1/69}

\bibitem[{{Girart} {et~al.}(2006){Girart}, {Rao}, \& {Marrone}}]{Girart2006}
{Girart}, J.~M., {Rao}, R., \& {Marrone}, D.~P. 2006, Science, 313, 812,
  \dodoi{10.1126/science.1129093}

\bibitem[{{Hennebelle} \& {Ciardi}(2009)}]{Hennebelle2009}
{Hennebelle}, P., \& {Ciardi}, A. 2009, \aap, 506, L29,
  \dodoi{10.1051/0004-6361/200913008}

\bibitem[{{Hennebelle} \& {Fromang}(2008)}]{Hennebelle2008a}
{Hennebelle}, P., \& {Fromang}, S. 2008, \aap, 477, 9,
  \dodoi{10.1051/0004-6361:20078309}

\bibitem[{{Hennebelle} \& {Inutsuka}(2019)}]{Hennebelle2019magnetic}
{Hennebelle}, P., \& {Inutsuka}, S.-i. 2019, Frontiers in Astronomy and Space
  Sciences, 6, 5, \dodoi{10.3389/fspas.2019.00005}

\bibitem[{{Hennebelle} \& {Teyssier}(2008)}]{Hennebelle2008b}
{Hennebelle}, P., \& {Teyssier}, R. 2008, \aap, 477, 25,
  \dodoi{10.1051/0004-6361:20078310}

\bibitem[{{Houde} {et~al.}(2016){Houde}, {Hull}, {Plambeck}, {Vaillancourt}, \&
  {Hildebrand}}]{Houde2016IV}
{Houde}, M., {Hull}, C. L.~H., {Plambeck}, R.~L., {Vaillancourt}, J.~E., \&
  {Hildebrand}, R.~H. 2016, \apj, 820, 38, \dodoi{10.3847/0004-637X/820/1/38}

\bibitem[{{Houde} {et~al.}(2009){Houde}, {Vaillancourt}, {Hildebrand},
  {Chitsazzadeh}, \& {Kirby}}]{Houde2009II}
{Houde}, M., {Vaillancourt}, J.~E., {Hildebrand}, R.~H., {Chitsazzadeh}, S., \&
  {Kirby}, L. 2009, \apj, 706, 1504, \dodoi{10.1088/0004-637X/706/2/1504}

\bibitem[{{Huang} {et~al.}(2024){Huang}, {Girart}, {Stephens}, {Fern{\'a}ndez
  L{\'o}pez}, {Arce}, {Carpenter}, {Cortes}, {Cox}, {Friesen}, {Le Gouellec},
  {Hull}, {Karnath}, {Kwon}, {Li}, {Looney}, {Megeath}, {Myers}, {Murillo},
  {Pineda}, {Sadavoy}, {S{\'a}nchez-Monge}, {Sanhueza}, {Tobin}, {Zhang},
  {Jackson}, \& {Segura-Cox}}]{Huang2024}
{Huang}, B., {Girart}, J.~M., {Stephens}, I.~W., {et~al.} 2024, \apjl, 963,
  L31, \dodoi{10.3847/2041-8213/ad27d4}

\bibitem[{{Huang} {et~al.}(2025{\natexlab{a}}){Huang}, {Girart}, {Stephens},
  {Fern{\'a}ndez L{\'o}pez}, {Myers}, {Zhang}, {Tobin}, {Cortes}, {Murillo},
  {Sadavoy}, {Arce}, {Carpenter}, {Kwon}, {Le Gouellec}, {Li}, {Looney},
  {Megeath}, {Cox}, {Karnath}, \& {Segura-Cox}}]{huang2025a}
---. 2025{\natexlab{a}}, \apj, 981, 30, \dodoi{10.3847/1538-4357/ad9ea4}

\bibitem[{{Huang} {et~al.}(2025{\natexlab{b}}){Huang}, {Girart}, {Stephens},
  {Myers}, {Zhang}, {Cortes}, {S{\'a}nchez-Monge}, {Fern{\'a}ndez L{\'o}pez},
  {Le Gouellec}, {Megeath}, {Murillo}, {Carpenter}, {Li}, {Liu}, {Looney},
  {Sadavoy}, {Karnath}, \& {Kwon}}]{huang2025b}
---. 2025{\natexlab{b}}, \apj, 984, 29, \dodoi{10.3847/1538-4357/adc30b}

\bibitem[{{Hull} \& {Plambeck}(2015)}]{hull2015debias}
{Hull}, C. L.~H., \& {Plambeck}, R.~L. 2015, Journal of Astronomical
  Instrumentation, 4, 1550005, \dodoi{10.1142/S2251171715500051}

\bibitem[{{Hull} {et~al.}(2017){Hull}, {Mocz}, {Burkhart}, {Goodman}, {Girart},
  {Cort{\'e}s}, {Hernquist}, {Springel}, {Li}, \& {Lai}}]{hull2017}
{Hull}, C. L.~H., {Mocz}, P., {Burkhart}, B., {et~al.} 2017, \apjl, 842, L9,
  \dodoi{10.3847/2041-8213/aa71b7}

\bibitem[{{Jiao} {et~al.}(2024){Jiao}, {Wang}, {Xu}, {Wang}, \&
  {Beuther}}]{jiao2024structure}
{Jiao}, W., {Wang}, K., {Xu}, F., {Wang}, C., \& {Beuther}, H. 2024, \aap, 686,
  A202, \dodoi{10.1051/0004-6361/202449182}

\bibitem[{{Joos} {et~al.}(2012){Joos}, {Hennebelle}, \& {Ciardi}}]{Joos2012}
{Joos}, M., {Hennebelle}, P., \& {Ciardi}, A. 2012, \aap, 543, A128,
  \dodoi{10.1051/0004-6361/201118730}

\bibitem[{{J{\o}rgensen} {et~al.}(2015){J{\o}rgensen}, {Visser}, {Williams}, \&
  {Bergin}}]{jorgensen2015molecule}
{J{\o}rgensen}, J.~K., {Visser}, R., {Williams}, J.~P., \& {Bergin}, E.~A.
  2015, \aap, 579, A23, \dodoi{10.1051/0004-6361/201425317}

\bibitem[{{Kataoka} {et~al.}(2015){Kataoka}, {Muto}, {Momose}, {Tsukagoshi},
  {Fukagawa}, {Shibai}, {Hanawa}, {Murakawa}, \& {Dullemond}}]{Kataoka2015}
{Kataoka}, A., {Muto}, T., {Momose}, M., {et~al.} 2015, \apj, 809, 78,
  \dodoi{10.1088/0004-637X/809/1/78}

\bibitem[{{Kauffmann} {et~al.}(2008){Kauffmann}, {Bertoldi}, {Bourke}, {Evans},
  \& {Lee}}]{kauffmann2008mambo}
{Kauffmann}, J., {Bertoldi}, F., {Bourke}, T.~L., {Evans}, N.~J., I., \& {Lee},
  C.~W. 2008, \aap, 487, 993, \dodoi{10.1051/0004-6361:200809481}

\bibitem[{{Krasnopolsky} \& {K{\"o}nigl}(2002)}]{Krasnopolsky2002}
{Krasnopolsky}, R., \& {K{\"o}nigl}, A. 2002, \apj, 580, 987,
  \dodoi{10.1086/343890}

\bibitem[{{Krasnopolsky} {et~al.}(2011){Krasnopolsky}, {Li}, \&
  {Shang}}]{Krasnopolsky2011}
{Krasnopolsky}, R., {Li}, Z.-Y., \& {Shang}, H. 2011, \apj, 733, 54,
  \dodoi{10.1088/0004-637X/733/1/54}

\bibitem[{{Kwon} {et~al.}(2022){Kwon}, {Pattle}, {Sadavoy}, {Hull},
  {Johnstone}, {Ward-Thompson}, {Di Francesco}, {Koch}, {Furuya}, {Doi}, {Le
  Gouellec}, {Hwang}, {Lyo}, {Soam}, {Tang}, {Hoang}, {Kirchschlager},
  {Eswaraiah}, {Fanciullo}, {Kim}, {Onaka}, {K{\"o}nyves}, {Kang}, {Lee},
  {Tamura}, {Bastien}, {Hasegawa}, {Lai}, {Qiu}, {Berry}, {Arzoumanian},
  {Bourke}, {Byun}, {Chen}, {Chen}, {Chen}, {Chen}, {Ching}, {Cho}, {Choi},
  {Choi}, {Chrysostomou}, {Chung}, {Coud{\'e}}, {Dai}, {Diep}, {Duan}, {Duan},
  {Eden}, {Fiege}, {Fissel}, {Franzmann}, {Friberg}, {Friesen}, {Fuller},
  {Gledhill}, {Graves}, {Greaves}, {Griffin}, {Gu}, {Han}, {Hatchell},
  {Hayashi}, {Houde}, {Inoue}, {Inutsuka}, {Iwasaki}, {Jeong}, {Kang},
  {Karoly}, {Kataoka}, {Kawabata}, {Kemper}, {Kim}, {Kim}, {Kim}, {Kim}, {Kim},
  {Kirk}, {Kobayashi}, {Kusune}, {Kwon}, {Lacaille}, {Law}, {Lee}, {Lee},
  {Lee}, {Lee}, {Lee}, {Li}, {Li}, {Li}, {Lin}, {Liu}, {Liu}, {Liu}, {Liu},
  {Lu}, {Mairs}, {Matsumura}, {Matthews}, {Moriarty-Schieven}, {Nagata},
  {Nakamura}, {Nakanishi}, {Ngoc}, {Ohashi}, {Park}, {Parsons}, {Peretto},
  {Priestley}, {Pyo}, {Qian}, {Rao}, {Rawlings}, {Rawlings}, {Retter},
  {Richer}, {Rigby}, {Saito}, {Savini}, {Seta}, {Shimajiri}, {Shinnaga},
  {Tahani}, {Tang}, {Tomisaka}, {Tram}, {Tsukamoto}, {Viti}, {Wang}, {Wang},
  {Whitworth}, {Wu}, {Xie}, {Yen}, {Yoo}, {Yuan}, {Yun}, {Zenko}, {Zhang},
  {Zhang}, {Zhang}, {Zhou}, {Zhu}, {de Looze}, {Andr{\'e}}, {Dowell}, {Eyres},
  {Falle}, {Robitaille}, \& {Loo}}]{kwon2022hro}
{Kwon}, W., {Pattle}, K., {Sadavoy}, S., {et~al.} 2022, \apj, 926, 163,
  \dodoi{10.3847/1538-4357/ac4bbe}

\bibitem[{Lazarian(2007)}]{Lazarian2007}
Lazarian, A. 2007, Journal of Quantitative Spectroscopy and Radiative Transfer,
  106, 225–256, \dodoi{10.1016/j.jqsrt.2007.01.038}

\bibitem[{{Lazarian} \& {Vishniac}(1999)}]{Lazarian1999}
{Lazarian}, A., \& {Vishniac}, E.~T. 1999, \apj, 517, 700,
  \dodoi{10.1086/307233}

\bibitem[{{Le Gouellec} {et~al.}(2019{\natexlab{a}}){Le Gouellec}, {Hull},
  {Maury}, {Girart}, {Tychoniec}, {Kristensen}, {Li}, {Louvet}, {Cortes}, \&
  {Rao}}]{LeGouellec2019}
{Le Gouellec}, V. J.~M., {Hull}, C. L.~H., {Maury}, A.~J., {et~al.}
  2019{\natexlab{a}}, \apj, 885, 106, \dodoi{10.3847/1538-4357/ab43c2}

\bibitem[{{Le Gouellec} {et~al.}(2019{\natexlab{b}}){Le Gouellec}, {Hull},
  {Maury}, {Girart}, {Tychoniec}, {Kristensen}, {Li}, {Louvet}, {Cortes}, \&
  {Rao}}]{le2019b}
---. 2019{\natexlab{b}}, \apj, 885, 106, \dodoi{10.3847/1538-4357/ab43c2}

\bibitem[{{Le Gouellec} {et~al.}(2020){Le Gouellec}, {Maury}, {Guillet},
  {Hull}, {Girart}, {Verliat}, {Mignon-Risse}, {Valdivia}, {Hennebelle},
  {Gonz{\'a}lez}, \& {Louvet}}]{le2020IMS}
{Le Gouellec}, V.~J.~M., {Maury}, A.~J., {Guillet}, V., {et~al.} 2020, \aap,
  644, A11, \dodoi{10.1051/0004-6361/202038404}

\bibitem[{{Li} {et~al.}(2013){Li}, {Krasnopolsky}, \& {Shang}}]{Li2013}
{Li}, Z.-Y., {Krasnopolsky}, R., \& {Shang}, H. 2013, \apj, 774, 82,
  \dodoi{10.1088/0004-637X/774/1/82}

\bibitem[{{Nagai} {et~al.}(1998){Nagai}, {Inutsuka}, \& {Miyama}}]{Nagai1998}
{Nagai}, T., {Inutsuka}, S.-i., \& {Miyama}, S.~M. 1998, \apj, 506, 306,
  \dodoi{10.1086/306249}

\bibitem[{{Nakamura} \& {Li}(2005)}]{Nakamura2005}
{Nakamura}, F., \& {Li}, Z.-Y. 2005, \apj, 631, 411, \dodoi{10.1086/432606}

\bibitem[{{Ossenkopf} \& {Henning}(1994)}]{ossenkopf1994dust}
{Ossenkopf}, V., \& {Henning}, T. 1994, \aap, 291, 943

\bibitem[{{Pattle} {et~al.}(2023){Pattle}, {Fissel}, {Tahani}, {Liu}, \&
  {Ntormousi}}]{Pattle2023}
{Pattle}, K., {Fissel}, L., {Tahani}, M., {Liu}, T., \& {Ntormousi}, E. 2023,
  in Astronomical Society of the Pacific Conference Series, Vol. 534,
  Protostars and Planets VII, ed. S.~{Inutsuka}, Y.~{Aikawa}, T.~{Muto},
  K.~{Tomida}, \& M.~{Tamura}, 193, \dodoi{10.48550/arXiv.2203.11179}

\bibitem[{{Pineda} {et~al.}(2020){Pineda}, {Segura-Cox}, {Caselli},
  {Cunningham}, {Zhao}, {Schmiedeke}, {Maureira}, \&
  {Neri}}]{pineda2020streamer}
{Pineda}, J.~E., {Segura-Cox}, D., {Caselli}, P., {et~al.} 2020, Nature
  Astronomy, 4, 1158, \dodoi{10.1038/s41550-020-1150-z}

\bibitem[{{Planck Collaboration} {et~al.}(2016{\natexlab{a}}){Planck
  Collaboration}, {Ade}, {Aghanim}, {Arnaud}, {Arroja}, {Ashdown}, {Aumont},
  {Baccigalupi}, {Ballardini}, {Banday}, {Barreiro}, {Bartolo}, {Battaner},
  {Benabed}, {Beno{\^\i}t}, {Benoit-L{\'e}vy}, {Bernard}, {Bersanelli},
  {Bielewicz}, {Bock}, {Bonaldi}, {Bonavera}, {Bond}, {Borrill}, {Bouchet},
  {Bucher}, {Burigana}, {Butler}, {Calabrese}, {Cardoso}, {Catalano},
  {Chamballu}, {Chiang}, {Chluba}, {Christensen}, {Church}, {Clements},
  {Colombi}, {Colombo}, {Combet}, {Couchot}, {Coulais}, {Crill}, {Curto},
  {Cuttaia}, {Danese}, {Davies}, {Davis}, {de Bernardis}, {de Rosa}, {de
  Zotti}, {Delabrouille}, {D{\'e}sert}, {Diego}, {Dolag}, {Dole}, {Donzelli},
  {Dor{\'e}}, {Douspis}, {Ducout}, {Dupac}, {Efstathiou}, {Elsner},
  {En{\ss}lin}, {Eriksen}, {Fergusson}, {Finelli}, {Florido}, {Forni},
  {Frailis}, {Fraisse}, {Franceschi}, {Frejsel}, {Galeotta}, {Galli}, {Ganga},
  {Giard}, {Giraud-H{\'e}raud}, {Gjerl{\o}w}, {Gonz{\'a}lez-Nuevo},
  {G{\'o}rski}, {Gratton}, {Gregorio}, {Gruppuso}, {Gudmundsson}, {Hansen},
  {Hanson}, {Harrison}, {Helou}, {Henrot-Versill{\'e}},
  {Hern{\'a}ndez-Monteagudo}, {Herranz}, {Hildebrandt}, {Hivon}, {Hobson},
  {Holmes}, {Hornstrup}, {Hovest}, {Huffenberger}, {Hurier}, {Jaffe}, {Jaffe},
  {Jones}, {Juvela}, {Keih{\"a}nen}, {Keskitalo}, {Kim}, {Kisner}, {Knoche},
  {Kunz}, {Kurki-Suonio}, {Lagache}, {L{\"a}hteenm{\"a}ki}, {Lamarre},
  {Lasenby}, {Lattanzi}, {Lawrence}, {Leahy}, {Leonardi}, {Lesgourgues},
  {Levrier}, {Liguori}, {Lilje}, {Linden-V{\o}rnle}, {L{\'o}pez-Caniego},
  {Lubin}, {Mac{\'\i}as-P{\'e}rez}, {Maggio}, {Maino}, {Mandolesi}, {Mangilli},
  {Maris}, {Martin}, {Mart{\'\i}nez-Gonz{\'a}lez}, {Masi}, {Matarrese},
  {McGehee}, {Meinhold}, {Melchiorri}, {Mendes}, {Mennella}, {Migliaccio},
  {Mitra}, {Miville-Desch{\^e}nes}, {Molinari}, {Moneti}, {Montier},
  {Morgante}, {Mortlock}, {Moss}, {Munshi}, {Murphy}, {Naselsky}, {Nati},
  {Natoli}, {Netterfield}, {N{\o}rgaard-Nielsen}, {Noviello}, {Novikov},
  {Novikov}, {Oppermann}, {Oxborrow}, {Paci}, {Pagano}, {Pajot}, {Paoletti},
  {Pasian}, {Patanchon}, {Perdereau}, {Perotto}, {Perrotta}, {Pettorino},
  {Piacentini}, {Piat}, {Pierpaoli}, {Pietrobon}, {Plaszczynski},
  {Pointecouteau}, {Polenta}, {Popa}, {Pratt}, {Pr{\'e}zeau}, {Prunet},
  {Puget}, {Rachen}, {Rebolo}, {Reinecke}, {Remazeilles}, {Renault}, {Renzi},
  {Ristorcelli}, {Rocha}, {Rosset}, {Rossetti}, {Roudier},
  {Rubi{\~n}o-Mart{\'\i}n}, {Ruiz-Granados}, {Rusholme}, {Sandri}, \&
  {Santos}}]{Planck2016XIX}
{Planck Collaboration}, {Ade}, P.~A.~R., {Aghanim}, N., {et~al.}
  2016{\natexlab{a}}, \aap, 594, A19, \dodoi{10.1051/0004-6361/201525821}

\bibitem[{{Planck Collaboration} {et~al.}(2016{\natexlab{b}}){Planck
  Collaboration}, {Ade}, {Aghanim}, {Alves}, {Arnaud}, {Arzoumanian},
  {Ashdown}, {Aumont}, {Baccigalupi}, {Banday}, {Barreiro}, {Bartolo},
  {Battaner}, {Benabed}, {Beno{\^\i}t}, {Benoit-L{\'e}vy}, {Bernard},
  {Bersanelli}, {Bielewicz}, {Bock}, {Bonavera}, {Bond}, {Borrill}, {Bouchet},
  {Boulanger}, {Bracco}, {Burigana}, {Calabrese}, {Cardoso}, {Catalano},
  {Chiang}, {Christensen}, {Colombo}, {Combet}, {Couchot}, {Crill}, {Curto},
  {Cuttaia}, {Danese}, {Davies}, {Davis}, {de Bernardis}, {de Rosa}, {de
  Zotti}, {Delabrouille}, {Dickinson}, {Diego}, {Dole}, {Donzelli}, {Dor{\'e}},
  {Douspis}, {Ducout}, {Dupac}, {Efstathiou}, {Elsner}, {En{\ss}lin},
  {Eriksen}, {Falceta-Gon{\c{c}}alves}, {Falgarone}, {Ferri{\`e}re}, {Finelli},
  {Forni}, {Frailis}, {Fraisse}, {Franceschi}, {Frejsel}, {Galeotta}, {Galli},
  {Ganga}, {Ghosh}, {Giard}, {Gjerl{\o}w}, {Gonz{\'a}lez-Nuevo}, {G{\'o}rski},
  {Gregorio}, {Gruppuso}, {Gudmundsson}, {Guillet}, {Harrison}, {Helou},
  {Hennebelle}, {Henrot-Versill{\'e}}, {Hern{\'a}ndez-Monteagudo}, {Herranz},
  {Hildebrandt}, {Hivon}, {Holmes}, {Hornstrup}, {Huffenberger}, {Hurier},
  {Jaffe}, {Jaffe}, {Jones}, {Juvela}, {Keih{\"a}nen}, {Keskitalo}, {Kisner},
  {Knoche}, {Kunz}, {Kurki-Suonio}, {Lagache}, {Lamarre}, {Lasenby},
  {Lattanzi}, {Lawrence}, {Leonardi}, {Levrier}, {Liguori}, {Lilje},
  {Linden-V{\o}rnle}, {L{\'o}pez-Caniego}, {Lubin}, {Mac{\'\i}as-P{\'e}rez},
  {Maino}, {Mandolesi}, {Mangilli}, {Maris}, {Martin},
  {Mart{\'\i}nez-Gonz{\'a}lez}, {Masi}, {Matarrese}, {Melchiorri}, {Mendes},
  {Mennella}, {Migliaccio}, {Miville-Desch{\^e}nes}, {Moneti}, {Montier},
  {Morgante}, {Mortlock}, {Munshi}, {Murphy}, {Naselsky}, {Nati},
  {Netterfield}, {Noviello}, {Novikov}, {Novikov}, {Oppermann}, {Oxborrow},
  {Pagano}, {Pajot}, {Paladini}, {Paoletti}, {Pasian}, {Perotto}, {Pettorino},
  {Piacentini}, {Piat}, {Pierpaoli}, {Pietrobon}, {Plaszczynski},
  {Pointecouteau}, {Polenta}, {Ponthieu}, {Pratt}, {Prunet}, {Puget}, {Rachen},
  {Reinecke}, {Remazeilles}, {Renault}, {Renzi}, {Ristorcelli}, {Rocha},
  {Rossetti}, {Roudier}, {Rubi{\~n}o-Mart{\'\i}n}, {Rusholme}, {Sandri},
  {Santos}, {Savelainen}, {Savini}, {Scott}, {Soler}, {Stolyarov}, {Sudiwala},
  {Sutton}, {Suur-Uski}, {Sygnet}, {Tauber}, {Terenzi}, {Toffolatti}, {Tomasi},
  {Tristram}, {Tucci}, {Umana}, {Valenziano}, {Valiviita}, {Van Tent},
  {Vielva}, {Villa}, {Wade}, {Wandelt}, {Wehus}, {Ysard}, {Yvon}, \&
  {Zonca}}]{planck2016XXXV}
---. 2016{\natexlab{b}}, \aap, 586, A138, \dodoi{10.1051/0004-6361/201525896}

\bibitem[{{Qiu} {et~al.}(2014){Qiu}, {Zhang}, {Menten}, {Liu}, {Tang}, \&
  {Girart}}]{Qiu2014}
{Qiu}, K., {Zhang}, Q., {Menten}, K.~M., {et~al.} 2014, \apjl, 794, L18,
  \dodoi{10.1088/2041-8205/794/1/L18}

\bibitem[{{Santos-Lima} {et~al.}(2012){Santos-Lima}, {de Gouveia Dal Pino}, \&
  {Lazarian}}]{SantosLima2012}
{Santos-Lima}, R., {de Gouveia Dal Pino}, E.~M., \& {Lazarian}, A. 2012, \apj,
  747, 21, \dodoi{10.1088/0004-637X/747/1/21}

\bibitem[{{Sokolov} {et~al.}(2019){Sokolov}, {Wang}, {Pineda}, {Caselli},
  {Henshaw}, {Barnes}, {Tan}, {Fontani}, \&
  {Jim{\'e}nez-Serra}}]{sokolov2019structure}
{Sokolov}, V., {Wang}, K., {Pineda}, J.~E., {et~al.} 2019, \apj, 872, 30,
  \dodoi{10.3847/1538-4357/aafaff}

\bibitem[{{Soler} {et~al.}(2013){Soler}, {Hennebelle}, {Martin},
  {Miville-Desch{\^e}nes}, {Netterfield}, \& {Fissel}}]{Soler2013}
{Soler}, J.~D., {Hennebelle}, P., {Martin}, P.~G., {et~al.} 2013, \apj, 774,
  128, \dodoi{10.1088/0004-637X/774/2/128}

\bibitem[{{Soler} {et~al.}(2017){Soler}, {Ade}, {Angil{\`e}}, {Ashton},
  {Benton}, {Devlin}, {Dober}, {Fissel}, {Fukui}, {Galitzki}, {Gandilo},
  {Hennebelle}, {Klein}, {Li}, {Korotkov}, {Martin}, {Matthews}, {Moncelsi},
  {Netterfield}, {Novak}, {Pascale}, {Poidevin}, {Santos}, {Savini}, {Scott},
  {Shariff}, {Thomas}, {Tucker}, {Tucker}, \& {Ward-Thompson}}]{Soler2017}
{Soler}, J.~D., {Ade}, P.~A.~R., {Angil{\`e}}, F.~E., {et~al.} 2017, \aap, 603,
  A64, \dodoi{10.1051/0004-6361/201730608}

\bibitem[{{Stephens} {et~al.}(2013){Stephens}, {Looney}, {Kwon}, {Hull},
  {Plambeck}, {Crutcher}, {Chapman}, {Novak}, {Davidson}, {Vaillancourt},
  {Shinnaga}, \& {Matthews}}]{Stephens2013}
{Stephens}, I.~W., {Looney}, L.~W., {Kwon}, W., {et~al.} 2013, \apjl, 769, L15,
  \dodoi{10.1088/2041-8205/769/1/L15}

\bibitem[{{Stephens} {et~al.}(2017){Stephens}, {Yang}, {Li}, {Looney},
  {Kataoka}, {Kwon}, {Fern{\'a}ndez-L{\'o}pez}, {Hull}, {Hughes}, {Segura-Cox},
  {Mundy}, {Crutcher}, \& {Rao}}]{Stephens2017}
{Stephens}, I.~W., {Yang}, H., {Li}, Z.-Y., {et~al.} 2017, \apj, 851, 55,
  \dodoi{10.3847/1538-4357/aa998b}

\bibitem[{{Stephens} {et~al.}(2023){Stephens}, {Lin},
  {Fern{\'a}ndez-L{\'o}pez}, {Li}, {Looney}, {Yang}, {Harrison}, {Kataoka},
  {Carrasco-Gonzalez}, {Okuzumi}, \& {Tazaki}}]{Stephens2023Nature}
{Stephens}, I.~W., {Lin}, Z.-Y.~D., {Fern{\'a}ndez-L{\'o}pez}, M., {et~al.}
  2023, \nat, 623, 705, \dodoi{10.1038/s41586-023-06648-7}

\bibitem[{{Tobin} {et~al.}(2013){Tobin}, {Bergin}, {Hartmann}, {Lee}, {Maret},
  {Myers}, {Looney}, {Chiang}, \& {Friesen}}]{tobin2013resolved}
{Tobin}, J.~J., {Bergin}, E.~A., {Hartmann}, L., {et~al.} 2013, \apj, 765, 18,
  \dodoi{10.1088/0004-637X/765/1/18}

\bibitem[{{Tobin} {et~al.}(2020){Tobin}, {Sheehan}, {Megeath},
  {D{\'\i}az-Rodr{\'\i}guez}, {Offner}, {Murillo}, {van 't Hoff}, {van
  Dishoeck}, {Osorio}, {Anglada}, {Furlan}, {Stutz}, {Reynolds}, {Karnath},
  {Fischer}, {Persson}, {Looney}, {Li}, {Stephens}, {Chandler}, {Cox},
  {Dunham}, {Tychoniec}, {Kama}, {Kratter}, {Kounkel}, {Mazur}, {Maud},
  {Patel}, {Perez}, {Sadavoy}, {Segura-Cox}, {Sharma}, {Stephenson}, {Watson},
  \& {Wyrowski}}]{tobin2020vla}
{Tobin}, J.~J., {Sheehan}, P.~D., {Megeath}, S.~T., {et~al.} 2020, \apj, 890,
  130, \dodoi{10.3847/1538-4357/ab6f64}

\bibitem[{{Van Loo} {et~al.}(2014){Van Loo}, {Keto}, \& {Zhang}}]{VanLoo2014}
{Van Loo}, S., {Keto}, E., \& {Zhang}, Q. 2014, \apj, 789, 37,
  \dodoi{10.1088/0004-637X/789/1/37}

\bibitem[{{Wang} {et~al.}(2022){Wang}, {V{\"a}is{\"a}l{\"a}}, {Shang},
  {Krasnopolsky}, {Li}, {Lam}, \& {Yuan}}]{Wang2022}
{Wang}, W., {V{\"a}is{\"a}l{\"a}}, M.~S., {Shang}, H., {et~al.} 2022, \apj,
  928, 85, \dodoi{10.3847/1538-4357/ac4d2e}

\bibitem[{{Whitney} {et~al.}(2003){Whitney}, {Wood}, {Bjorkman}, \&
  {Wolff}}]{whitney2003radiative}
{Whitney}, B.~A., {Wood}, K., {Bjorkman}, J.~E., \& {Wolff}, M.~J. 2003, \apj,
  591, 1049, \dodoi{10.1086/375415}

\bibitem[{{Yang} {et~al.}(2016){Yang}, {Li}, {Looney}, {Cox}, {Tobin},
  {Stephens}, {Segura-Cox}, \& {Harris}}]{Yang2016}
{Yang}, H., {Li}, Z.-Y., {Looney}, L.~W., {et~al.} 2016, \mnras, 460, 4109,
  \dodoi{10.1093/mnras/stw1253}

\bibitem[{{Zhang} {et~al.}(2014){Zhang}, {Qiu}, {Girart}, {Liu}, {Tang},
  {Koch}, {Li}, {Keto}, {Ho}, {Rao}, {Lai}, {Ching}, {Frau}, {Chen}, {Li},
  {Padovani}, {Bontemps}, {Csengeri}, \& {Ju{\'a}rez}}]{Zhang2014}
{Zhang}, Q., {Qiu}, K., {Girart}, J.~M., {et~al.} 2014, \apj, 792, 116,
  \dodoi{10.1088/0004-637X/792/2/116}

\end{thebibliography}
\end{document}